\documentclass[a4paper,11pt]{article}
\pdfoutput=1 

\usepackage{color} 
\usepackage{xcolor} 
\usepackage{graphicx} 
\usepackage{afterpage} 
\usepackage{stmaryrd}
\usepackage[nottoc]{tocbibind} 
\usepackage{chngcntr}
\usepackage{fnpct}
\usepackage{subcaption} 
\usepackage{tikz} 
\usetikzlibrary{shapes,arrows,positioning,automata,backgrounds,calc,er,patterns,decorations.markings, shapes.misc} 
\usepackage[compat=1.1.0]{tikz-feynman} 
\usepackage{circuitikz} 
\ctikzset{bipoles/length=.8cm}
\usepackage{physics} 
\usepackage[bottom, marginal, perpage]{footmisc} 
\usepackage[nice]{nicefrac} 
\usepackage{units} 
\usepackage{mathtools} 
\usepackage{wrapfig} 

\usepackage{fancyhdr}
\usepackage{caption}
\usepackage{tabularx}
\usepackage{booktabs}
\usepackage{multirow}
\usepackage{multicol}
\usepackage{geometry}
\usepackage{setspace}
\usepackage{textcase}
\usepackage[titles]{tocloft}
\usepackage{etoolbox}
\newcommand{\degree}{^\circ}

\usepackage{jinstpub} 
\usepackage{cleveref}
\usepackage{lineno}

\definecolor{schmucki}{RGB}{153, 0, 0}
\definecolor{felix}{RGB}{0, 101, 189}
\definecolor{christian}{RGB}{41, 49, 123}

\newcommand*{\invisiblepar}{{\setlength{\parfillskip}{0pt}\par}\vskip-\parskip\noindent\ignorespaces}

\title{A self-monitoring precision calibration light source for large-volume neutrino telescopes}

\author[a,b,1]{F.~Henningsen,\note{Corresponding author.}}
\author[a]{M.~Boehmer,}
\author[a,d]{A. G\"artner,}
\author[a]{L.~Geilen,}
\author[a]{R.~Gernhäuser,}
\author[c]{H.~Heggen,}
\author[a]{K.~Holzapfel,}
\author[a]{C.~Fruck,}
\author[a]{L. Papp,}
\author[a]{I. C. Rea,}
\author[a]{E.~Resconi,}
\author[a]{F.~Schmuckermaier,}
\author[a]{C.~Spannfellner,}
\author[c]{M.~Traxler.}

\affiliation[a]{Physik-Department, Technische Universit\"at M\"unchen, D-85748 Garching, Germany}
\affiliation[b]{Max-Planck-Insitut f\"ur Physik, D-80805 Munich, Germany}
\affiliation[c]{GSI Helmholtz Centre for Heavy Ion Research GmbH, Darmstadt, Germany}
\affiliation[d]{Department of Physics, University of Alberta, Edmonton, Alberta, Canada T6G 2E1}

\emailAdd{felix.henningsen@tum.de}

\abstract{With the rise of neutrino astronomy using large-volume detector arrays, calibration improvements of optical media and photosensors have emerged as significant means to reduce detector systematics. To improve understanding of the detector volume and its instrumentation, we developed an absolutely-calibrated, self-monitoring, isotropic, nanosecond, high-intensity calibration light source called "Precision Optical Calibration Module" (POCAM). This, now third iteration, of the instrument was developed for an application in the IceCube Upgrade but, with a modular instrument communications and synchronization backend, can provide a calibration light source standard for any large-volume photodetector array. This work summarizes the functional principle of the POCAM and all related device characteristics as well as its precision calibration procedure. The latter provides fingerprint-characterized instruments with knowledge on absolute and relative behavior of the emitted light pulses as well as their temperature dependencies.}


\begin{document}
\maketitle
\flushbottom

\section{Large-volume detectors \& calibration}
\label{sec:telescopes}

\subsection{Detector arrays}
\label{subsec:array}
Neutrinos have proven to be highly valuable cosmic probes for fundamental physics research and the newly emerging field of multi-messenger astronomy. Extremely light and electrically neutral leptons only interact by gravitation and the weak force and can therefore travel long distances through the cosmos without attenuation. Unlike cosmic rays, which are electrically charged and can be deflected by magnetic fields, neutrinos point back to their origin. Sources accelerating cosmic rays are also believed to produce high energy neutrinos in an energy range between GeV to EeV \cite{2015RPPh...78l6901A}. Neutrinos are the unique interaction and decay products of hadronic cosmic rays and are thus one of the most promising routes to discover the origins of high-energy cosmic rays. At energies from MeV to GeV, neutrinos emitted from the sun and supernovae~\cite{2011A&A...535A.109A} offer insights into the evolution of stars, enable measurements of flavor oscillations~\cite{PhysRevLett.114.171101}, and with that establish evidence for physics beyond the Standard Model~\cite{Ahlers:2018mkf}.

To detect these elusive particles, scientists employ large-volume neutrino telescopes. These typically consist of a three-dimensional array of photomultiplier tubes (PMTs) surrounded by a transparent dielectric medium (e.g. water or ice). Neutrinos traversing the detector interact inside the detector medium with a small probability and produce highly energetic charged particles. If these charged particles are travelling fast enough through the detector medium, Cherenkov light is produced. Measurements of the Cherenkov radiation by the PMT array then allows the reconstruction of the neutrino's energy and direction \cite{reconstruction_icecube}. 

The primary high-energy neutrino interaction channel is deep-inelastic scattering with nuclei in the detector medium. In general, neutrinos can undergo a charged current (CC) interaction in the form of $\nu_l + X \rightarrow l^- + Y$ or a neutral current (NC) interaction in the form of $\nu_l + X \rightarrow \nu_l + Y $, where the same interaction is also possible with $\bar{\nu}_l$ and $l^+$ respectively and $l=e, \mu, \tau$. Depending on the type of interaction and flavor of the incoming neutrino, different event signatures arise in the detector. NC interactions of all flavors and CC interactions of electron neutrinos produce cascade events caused by the initial hadronic shower and in the latter case also the electromagnetic shower resulting from the outgoing electrons. Cascade events appear as isotropic point like sources of Cherenkov light. Due to the short distances traveled by leptons produced in cascade events, they deposit all of their emitted energy in the detector volume and allow for a comparatively accurate energy reconstruction, whereas the directional reconstruction is based on slight light intensity and timing asymmetries and is less accurate. This also makes the distinction between NC and CC interactions exceedingly difficult. In contrast,outgoing muons from muon neutrino CC interactions can travel for several kilometers through the detector medium, producing Cherenkov light on their way. These track events only make a rough energy reconstruction possible, since only a fraction of the whole track is contained in the detector volume. On the other hand, the elongated geometry enables a more precise directional reconstruction. CC interactions caused by tau neutrinos produce a cascade at the first interaction vertex, a track during the short propagation of the produced tau lepton and then a second cascade caused by the tau decay. The resulting signature is called 'double bang', but is nearly indistinguishable from regular cascade events due to the short distance of the two cascades. Most of the background in neutrino detectors is caused by muons and neutrinos originating from cosmic-ray air showers.
\begin{wrapfigure}{l}{0.5\textwidth} 
   \vspace{-5pt}
  \begin{center}
    \includegraphics[width=0.48\textwidth]{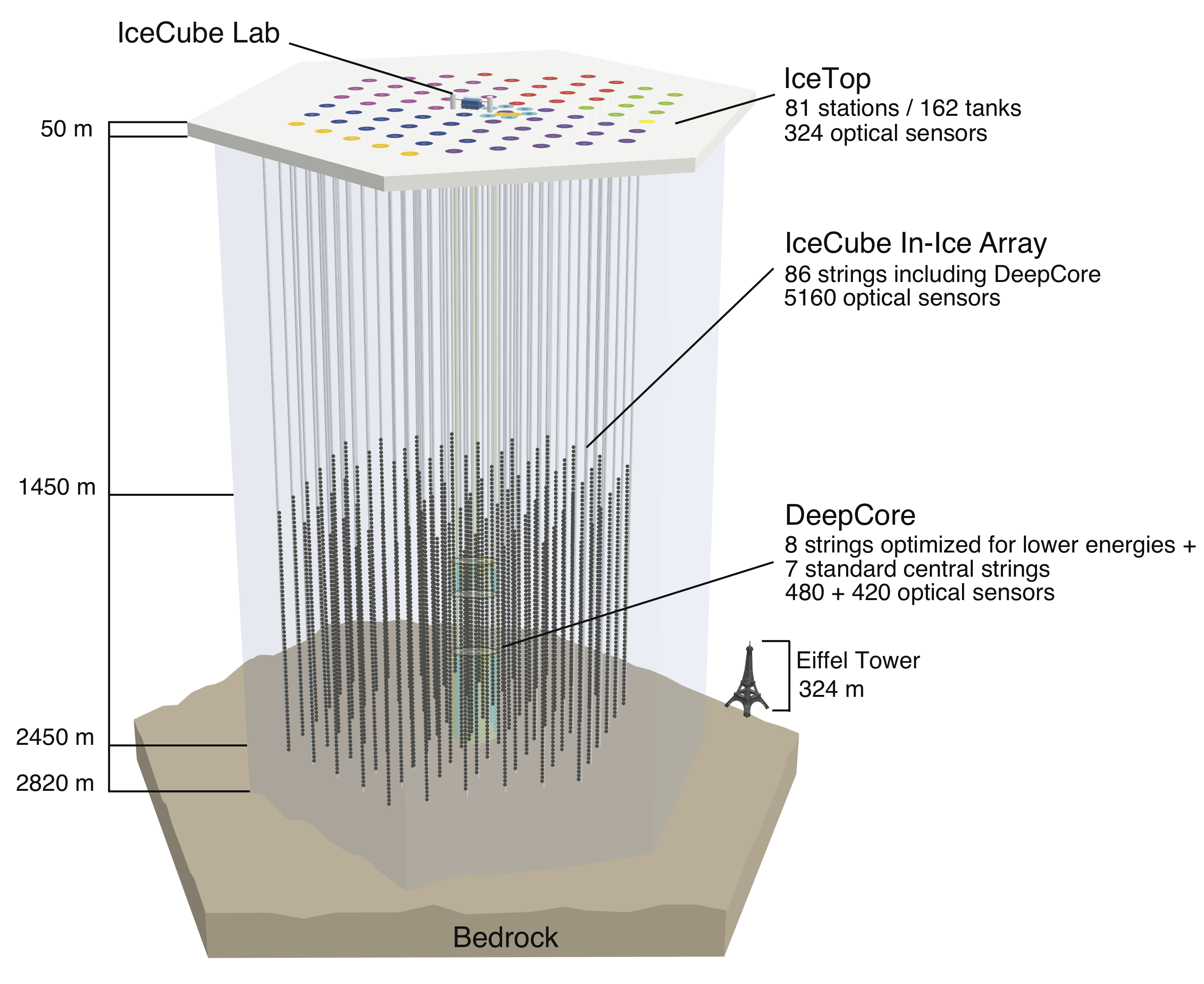}
  \end{center}
  \vspace{-15pt}
  \caption{The IceCube Neutrino Observatory with the in-ice array, its sub-array DeepCore, and the cosmic-ray air shower array IceTop \cite{Aartsen:2016nxy}.}
  \label{fig:icecube}
\end{wrapfigure}
 Neutrino detectors can use the Earth to shield most of the muon background when searching for neutrinos from the opposite hemisphere. However, the Earth increasingly absorbs neutrinos with energies above several tens of TeV, which means high-energy neutrinos must be observed by looking in the same hemisphere as the detector. To remove the large muon backgrounds from this direction, the analysis can focus exclusively on very high energies (above a few PeV) or select events with their primary vertex inside the detector. The latter approach naturally excludes background muons and can be used down to energies as low as $10\,$TeV. In addition the coincidence of a starting event and a muon can be used to reject atmospheric neutrinos origination from the same air shower as the muon~\cite{2013Sci...342E...1I}.

Since both the cross section of neutrino interactions and the flux of astrophysical neutrinos at Earth is small, large volumes of detector medium in the scale of km$^3$ are required to observe more than a handful of astrophysical events per year. Due to these requirements, PMT arrays are embedded in natural sites containing large volumes of glacial ice, lake water or sea water. The first-ever neutrino telescope in ice, AMANDA \cite{amanda}, was deployed between in 1993/94 at the geographic South Pole and was followed by several further upgrades and extensions. The development of drilling techniques, characterization of the optical properties of the glacial ice and improvement of reconstruction methods among others laid the groundwork for the major successor experiment: The IceCube Neutrino Observatory \cite{Aartsen:2016nxy}, the first cubic-kilometer neutrino detector. IceCube was completed in 2010 and its in-ice array consists of 5160 digital optical modules (DOMs) on 86 strings, deployed between a depth of 1450 m and 2450 m. Part of the in-ice array includes the DeepCore sub-array, which consists of 8 strings with closely-spaced DOMs to extend the sensitivity of IceCube down to 10 GeV~\cite{Collaboration:2011ym}.
\invisiblepar
\begin{wrapfigure}{r}{0.3\textwidth}
  \vspace{-15pt}
  \begin{center}
    \includegraphics[width=0.28\textwidth]{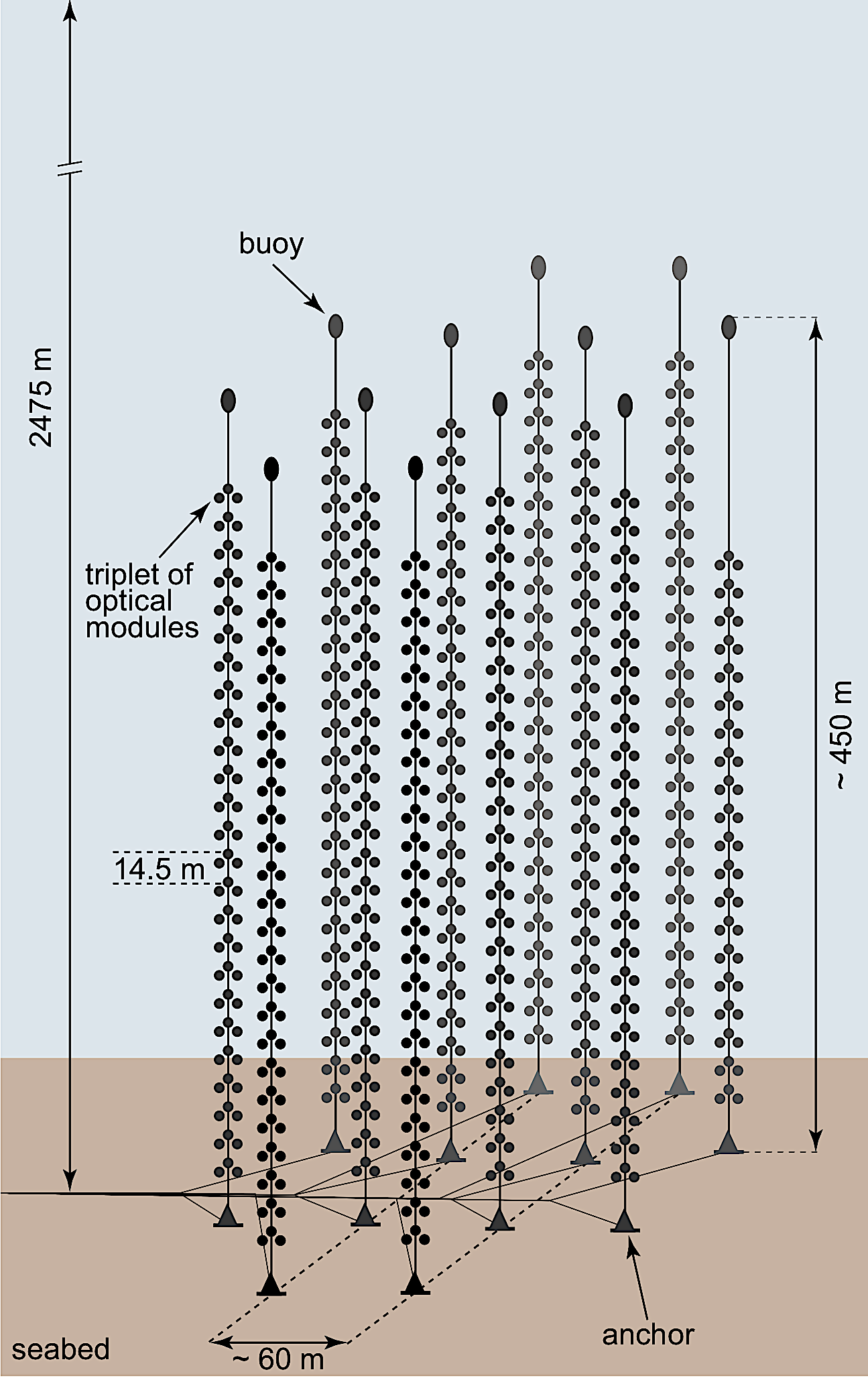}
  \end{center}
  \vspace{-10pt}
  \caption{Schematic view of the ANTARES detector \cite{AGERON201111}.}
  \label{fig:antares}
\end{wrapfigure}
The next stage of the IceCube project will be the deployment of the IceCube Upgrade \cite{Ishihara:2019aao} during the Antarctic Summer season 2022/23. Consisting of seven new strings, each carrying newly developed DOMs and calibration devices, the Upgrade will extend the detector's science capabilities in the low energy region and enable a re-calibration of the detector. New calibrations are made possible by deploying a variety of novel calibration devices, including the Precision Optical Calibration Module (POCAM) described here. The improved understanding of the optical properties of the South Pole ice and the detector response will result in an enhanced calibration of the detector and a re-analysis of archival data.

The first neutrino telescopes to use water as a detection medium were realized in the Mediterranean sea with the ANTARES detector~\cite{AGERON201111} and in Lake Baikal with the BAIKAL neutrino telescope NT200~\cite{2006NIMPA.567..433A}. These pearly experiments are being followed up with larger-scale detectors both in the Mediterranean Sea with the cubic-kilometer neutrino telescope (KM3NeT)~\cite{Margiotta:2014gza} as well as the Gigaton Volume Detector (GVD) in Lake Baikal~\cite{2019arXiv190805427B}. Plans for an additional cubic-kilometer scale detector in the Northern Pacific Ocean, the Pacific Ocean Neutrino Explorer (P-ONE), are currently ongoing and the pathfinder mission STRAW indicates this is a promising location~\cite{Bedard:2018zml}. The general geometries of neutrino interactions are similar for both water- and ice-based detectors except for differences in the optical properties and the systematic uncertainties faced by the different environments. Understanding the optical properties of the detector medium is crucial because the reconstruction of events is typically based on the intensity and arrival times of secondary photons which have undergone single or multiple scattering since the initial Cherenkov emission. The following section gives a qualitative summary of how systematic uncertainties can be estimated and improved with the aid of artificial calibration light sources.

\subsection{Systematic uncertainties}
\label{subsec:systematics}
The first main source of systematic uncertainties are the response and behavior of the optical modules. The latter typically consist of PMTs embedded in a gel-coated glass sphere. Both, the quantum efficiency of the PMT alone and the efficiency of a fully integrated module can be accurately determined in the laboratory. However, different conditions on the experiment site and ageing effects of the PMT components can cause deviations from the laboratory findings over time. Recurring \textit{in-situ} measurements of the detection efficiency are therefore advantageous. Furthermore, the saturation of the PMTs at high light levels can have an impact on event reconstruction. For a low number of arriving photons, the charge output of a PMT is proportional to the number of measured photons. For high photoelectron rates this linear behavior can break down and the PMT saturates. In case of the IceCube DOMs, this saturation limit is reached at around 31 photoelectrons per nanosecond \cite{2010NIMPA.618..139A}. For most neutrino events, the majority of PMTs will measure low light levels and their response will be linear. However, some high-energy neutrino interactions can deposit enough energy to saturate close-by PMTs. Absent a model of the nonlinear behavior of PMTs near their saturation limits, saturated PMTs must be excluded from event reconstruction.

A second major systematic uncertainty is the geometry of the detector array. In this regard, detectors embedded in glacial ice benefit from a static geometry. Once deployed, the positions of individual modules can be estimated from the deployment data and then be verified by trilateration. In water based detectors, the strings are flexible structures fixed at the seabed or lake bottom and held under tension by buoys. Hence, currents in the water are able to move and rotate the strings. Especially in sea water, these currents can change the geometry significantly and a frequent redetermination of the detector geometry via optical or acoustic systems is essential. 

The third major source of uncertainties are the optical properties of the detector medium \cite{2019PhRvD..99c2007A}. In general, propagation of light in transparent media is determined by the effects of scattering and absorption. The former results in a directional change, the latter in a loss of the photon. 
\invisiblepar
\begin{wrapfigure}{l}{0.45\textwidth}
  \vspace{-15pt}
  \begin{center}
    \includegraphics[width=0.43\textwidth]{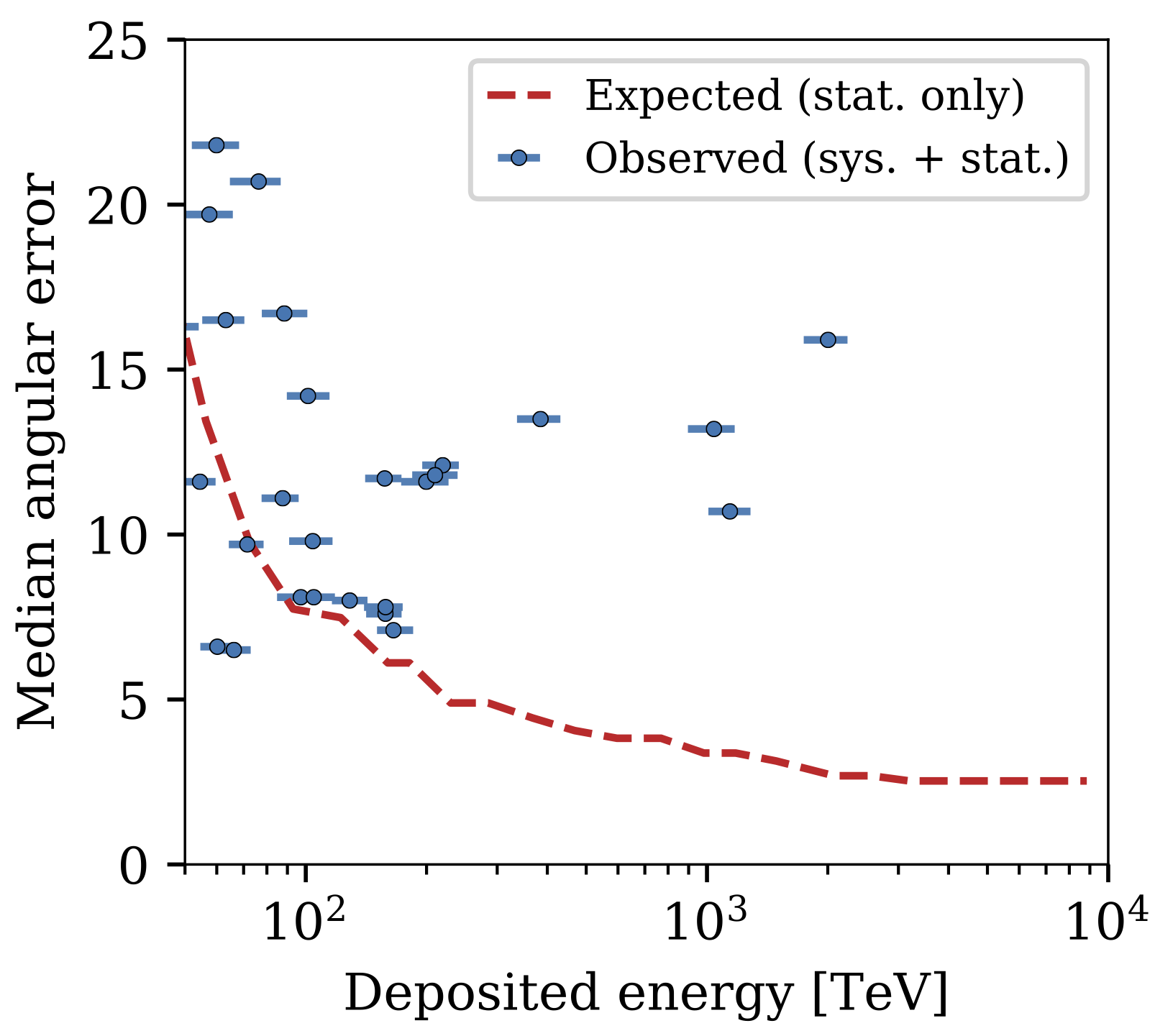}
  \end{center}
  \vspace{-10pt}
  \caption{Observed median angular error and statistical limit of fully contained high energy cascade directional reconstruction in IceCube as a function of reconstructed deposited energy \cite{Ishihara:2019aao}.}
  \label{fig:ang_err}
  \vspace{-5pt}
\end{wrapfigure}
The typical parametrization of these effects is achieved via the definition of a scattering length and an absorption length, which marks the distance after which the probability of the photon not being influenced by the respective effect drops to $1/e$. For detectors embedded in glacial ice, the scattering length is considerably shorter than the absorption length \cite{2013NIMPA.711...73A}, whereas for water-based detectors, absorption is the prevailing effect \cite{Bagley:2009wwa}. An accurate depth-dependent estimation of both scattering and absorption forms the basis of the optical characterization of the detector medium and, in the case of water based detectors, these properties need to be  monitored with \textit{in-situ} instruments frequently to account for changes in the environment. In the case of IceCube, an additional anisotropic scattering effect \cite{2013ICRC...33.3338C} has been observed. The optical anisotropy of the ice produces an azimuthal dependency in photon propagation. Furthermore, the refrozen ice around the DOMs in the drill hole, commonly referred to as 'Hole Ice' \cite{2013NIMPA.711...73A}, exhibits different optical behavior compared to bulk ice. The main difference is a much shorter scattering length, making it possible for photons approaching the back-side of the DOM to still be detected when scattered around the DOM towards the the PMT, and for photons heading towards the PMT-side of the DOM to be scattered away  more frequently. Together, these systematic uncertainties can significantly restrict the accuracy of event reconstruction, especially the directional reconstruction of high energy cascade events~\cite{Ishihara:2019aao}. Figure \ref{fig:ang_err} shows estimates of the angular uncertainty of high energy cascade events in IceCube as a function of reconstructed deposited energy. The dashed line represents the angular error with perfect knowledge of the detector response and optical properties of the ice, hence only the statistical uncertainty. The estimated total median angular errors, however, are far above the statistical limit, indicating a significant restriction on the reconstruction caused by a lack of knowledge about the contributing systematic uncertainties.

\subsection{Calibration with artificial light sources}
\label{subsec:calsources}
Artificial light sources are the most common approach to ensuring accurate characterization of detector systematic uncertainties. This section provides an overview of the contemporary techniques and methods used for the calibration of neutrino telescopes and discusses potential improvements, using the POCAM as an absolutely calibrated and isotropic light source. The main artificial light source present in IceCube is the LED flasher board \cite{Aartsen:2016nxy}. It contains 12 LEDs mounted circularly around the board covering six different azimuth angles (with $60^{\circ}$ spacing) and two different zenith angles. LEDs can be flashed with a rate of up to to 610$\,$Hz and emit a spectrum centered around 399$\,$nm, with an intensity range of $10^6 -10^{10}$ photons. 

The estimation of the bulk ice parameters were performed using data sets created by the LED flasher boards \cite{2013NIMPA.711...73A}. The ice is parameterized by six global parameters (see [14] for details), the depth-dependent temperature, and the scattering and absorption length at 400$\,$nm, averaged over layers of 10$\,$m thickness. A photon propagator was then used to simulate the propagation of light emitted by the flasher boards under varying ice parameters. Using a maximum likelihood analysis, a global fit for all parameters to the real flasher data can then be performed, which results in a table of estimates of depth-dependent values for the scattering and absorption length. For the detector volume below depths of 2100$\,$m, the resulting values of the absorption length are ranging from about 100$\,$m - 200$\,$m, while the scattering lengths range from around 30$\,$m - 80$\,$m, making scattering the dominant optical effect \cite{2013NIMPA.711...73A}. The same data set is also utilized to fit the parameterization of the observed anisotropy in the bulk ice \cite{2013ICRC...33.3338C}. The effect is currently characterized by three free parameters defining a diagonal matrix, which modifies the scattering behavior. However, the accuracy of the model is limited by a depth-dependence of the anisotropy and the rather uncertain emission profiles and pointings of the flasher LEDs. Due to the isotropic light emission of the POCAM, we expect that there is potential for an improvement of the anisotropy fit, since no initial light emission profile has to be assumed.

The hole ice effect is described as a modification of the angular acceptance curve of the DOM. Only for certain low energy event reconstructions, where the photon origin is close to the receiving DOM and no straight wavefront can be assumed, is the hole ice modeled as an actual ice column with optical properties deviating from the bulk ice~\cite{Aartsen_2019}. Since the POCAMs will be deployed at several depths, they offer the possibility of an improved \textit{in-situ} acceptance measurement affected by hole ice~\cite{Resconi:2017mad}. The optical efficiency model of the DOMs is supplemented after deployment with \textit{in-situ} measurements of down going, minimal ionizing muons, since they are abundant and have a fairly known light emission \cite{Aartsen:2016nxy}. The flasher board LEDs are not suitable here due to their unknown emission intensity. The absolution calibration of the POCAM means it is not affected by the normalization uncertainty of the LED flashers, and thus will provide an independent {\sl in situ} measurement of DOM efficiencies. Additionally, the POCAMs can contribute to the investigation of the PMT saturation. Precise knowledge about the intensity and an extensive dynamical range of the emitting light source enables an \textit{in-situ} characterization of the non-linear PMT response. This allows the inclusion of a higher number of otherwise neglected DOMs in high energy event reconstructions, which increases the overall detected charge and hence decreases the statistical uncertainty. Moreover, the efficiency and linearity behavior of modules can drift over time due to aging of the DOM hardware. A light source with absolute calibration is thus valuable for detecting long-term drifts in detector response.

Neutrino detectors deployed in lakes and seawater have developed a variety of similar calibration tools for studying the time-varying optical properties of the water. The KM3NeT detector will be located at several sites in the Mediterranean Sea. One of them, situated about 10$\,$km west of the ANTARES telescope, was already intensively studied by the ANTARES collaboration \cite{Aguilar:2004nw}. A dedicated detector line containing two separated spheres (light source and PMT) was deployed to measure absorption and scattering lengths with blue light ($\lambda$ = 473$\,$nm) and UV light ($\lambda$ = 375$\,$nm). The absorption length is around 60$\,$m for blue and 25$\,$m for UV light. In contrast, the scattering length is about 260$\,$m for blue and 120$\,$m for UV, illustrating an optical behavior opposite to glacial ice. Similar setups, developed by the NESTOR and NEMO collaborations, have been utilized to characterize the remaining KM3NeT sites \cite{Bagley:2009wwa}. Additional light sources for the calibration of the time offset between DOMs were implemented in ANTARES in the form of the Optical Beacon System \cite{2007NIMPA.578..498A}. It consists of pulsed LEDs and lasers located throughout the detector enabling a sub-nanosecond timing calibration. The developed system in KM3NeT, the Nanobeacon \cite{nanobeacon}, is composed of LEDs installed in the upper parts of each DOM. The optical water properties in Lake Baikal were determined with the aid of lasers \cite{Balkanov:1999uq} and later measurements with a deployed POCAM prototype confirmed the findings \cite{holzapfel:thesis:2019}.

It becomes apparent that any large-scale neutrino telescope has a comparable approach to its calibration and hence a similar demand in regards of calibration devices. The next generation of upcoming neutrino telescopes (Baikal-GVD, KM3NET, IceCube-Gen2 \& P-ONE) as well as currently operating telescopes benefit from precise and well-tested light sources. Especially for IceCube, significantly limiting systematic uncertainties indicate the need for an improved calibration, in order to fully exploit the detector's scientific capability. Due to its absolute intensity calibration, large dynamic range, short nanosecond pulses and self-monitoring ability, the POCAM has potential to function as a standard calibration light source for any large-scale optical neutrino telescopes.

\newpage

\section{Precision Optical Calibration Module}
\label{sec:pocam}
\begin{figure}[h!]
    \centering
    \begin{subfigure}[b]{0.225\textwidth}
    	\centering
    	\includegraphics[width=0.825\textwidth]{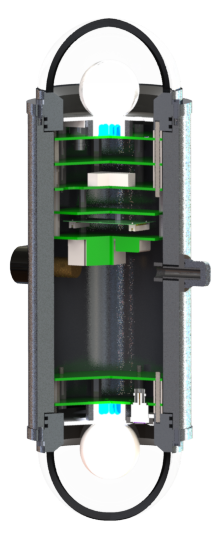}
    	\caption[]{Full instrument}
    	\label{fig:housing-all}
    \end{subfigure}
    \begin{subfigure}[b]{0.6725\textwidth}
    	\centering
    	\includegraphics[width=0.775\textwidth]{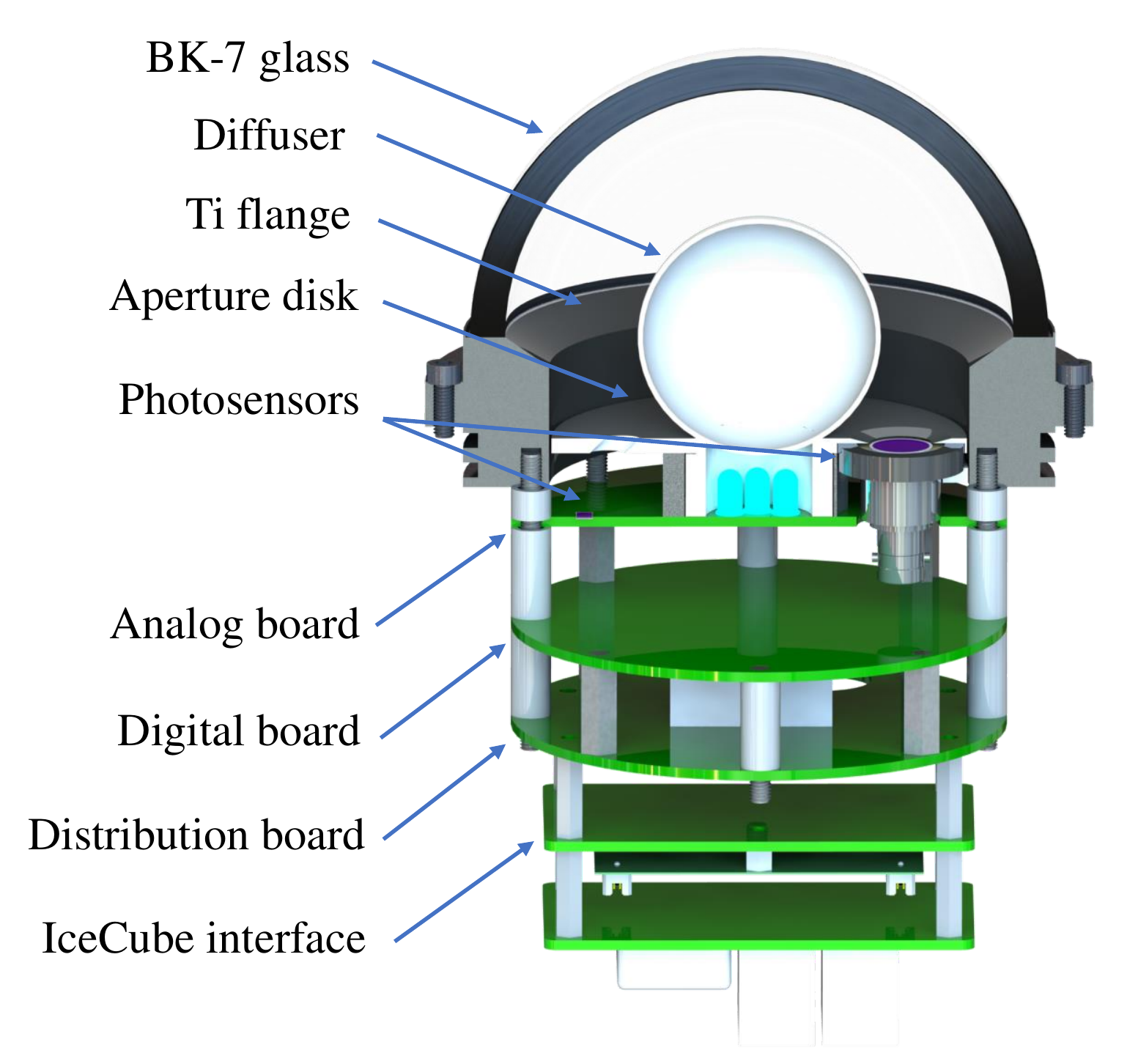}
    	\caption[]{Flange assembly cut-view with all sub-components.}
    	\label{fig:housing-flange}
    \end{subfigure}
    \caption{Full instrument (a) and flange assembly (b) cut-view of the POCAM instrument. In the flange assembly all the sub-components of one hemisphere are visible and annotated. For details on the latter refer to the text.}
    \label{fig:housing-cut}
\end{figure}

The need for higher precision calibration in neutrino telescope arrays has been evident for a number of years. In this scope, we developed a novel calibration light source providing isotropic self-monitored light pulses with large dynamic range~\cite{Bedard:2018zml,Resconi:2017mad,Fruck:2019vam}. This Precision Optical Calibration Module (POCAM) pursues the goal of providing a detector-independent calibration light source standard for large-volume photosensor arrays used for high- and low-energy neutrino detection. In general, this calibration instrument consists of diffuse isotropic light flashers, photosensors for self-monitoring as well as necessary electronics for readout and data aquisition (DAQ). All of these components are located in a pressure housing enabling deployment in deep water or other harsh environments. A cut-view of the POCAM is shown in \cref{fig:housing-all} with a detailed view of the flange assembly in \cref{fig:housing-flange}. The following subsections will summarize the development results of all device sub-components and will go into detail about their performance.
\par
\begin{figure}[h!]
	\centering
	\includegraphics[width=0.95\textwidth]{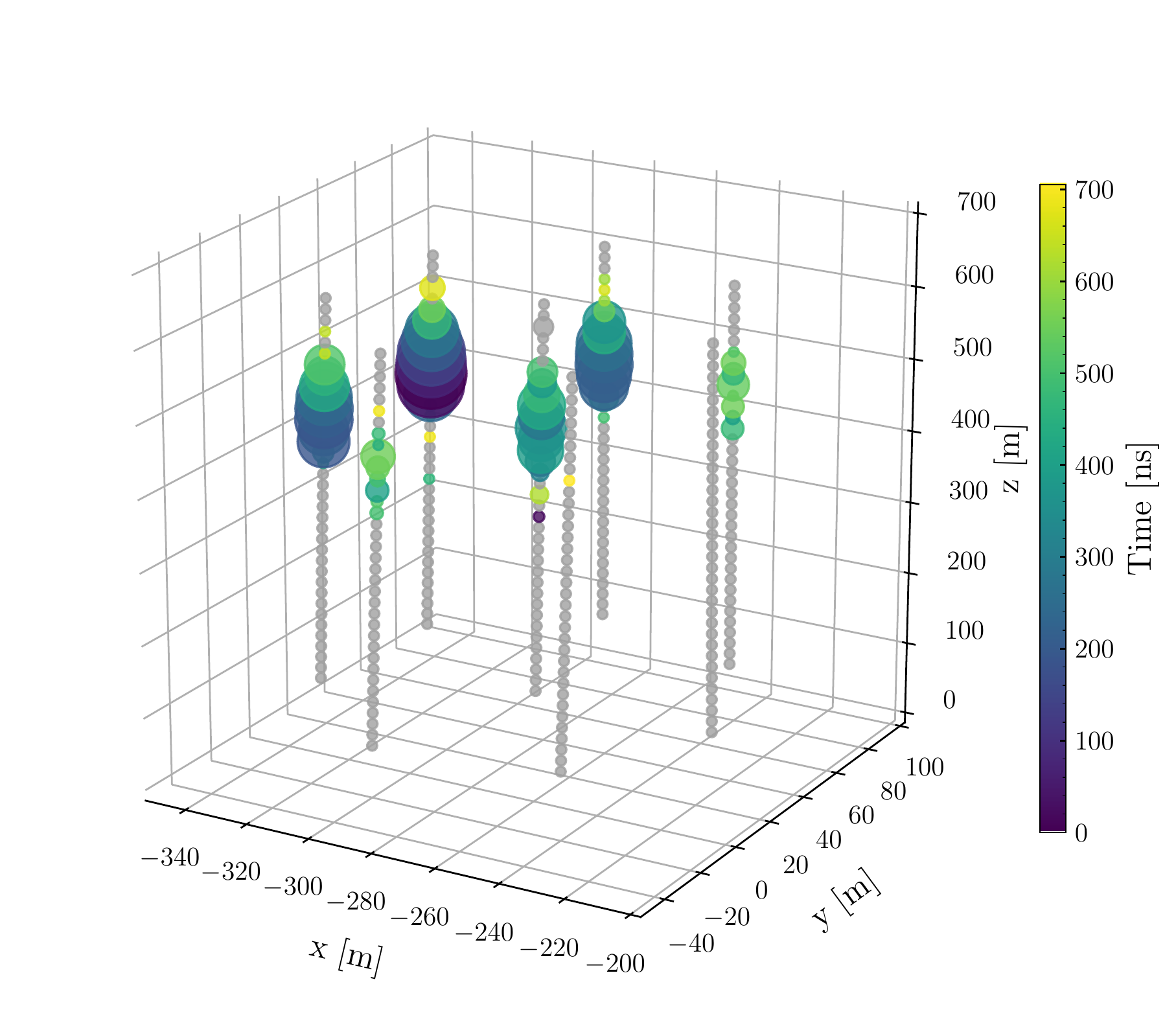}
	\caption[POCAM event view in GVD.]{First POCAM prototype event view within a GVD cluster. With size representing integrated charge and color arrival time of the light peak at observing, down-facing photosensors, the POCAM location on the back left string can be inferred by eye. Data provided by the GVD collaboration.}
	\label{fig:baikal-event}
\end{figure}

\noindent
The POCAM developments started in late 2014 with conceptual studies and simulations~\cite{Jurkovic:2016kxn,Ackermann:2017pja}. After around two years of investigating the baseline instrument performance requirements, a first prototype module was developed.
%
%
To provide proof-of-concept of the device, we collaborated with the GVD telescope located in Lake Baikal to deploy the first prototype in 2017. This first deployment has proven successful and the POCAM was able to reproduce~\cite{holzapfel:thesis:2019} the measured optical properties of the Baikal Lake water~\cite{Balkanov:2002ni,Balkanov:1999uq}. An event view of a POCAM light pulse recorded in the GVD cluster is shown in \cref{fig:baikal-event} and illustrates the detector coverage achieved with this instrument iteration. Following the successful application in Russia, three second-iteration instruments have been deployed in the STRAW experiment in 2018~\cite{Bedard:2018zml}. This detector uses the POCAM to probe the optical properties of the North-East Pacific deep sea and. During this deployment we were able to significantly extend and improve the POCAM performance and self-calibration. Further changes and improvements of the instrument where then done in the scope of the IceCube Upgrade developments~\cite{Ishihara:2019aao,Fruck:2019vam} during 2019. A total of 30 POCAM instruments will be produced with 21 being planned for the Upgrade detector volume~\cite{Fruck:2019vam}. This current instrument design baseline and its performance is the main focus of this work. 
\par\noindent
\subsection{Instrument housing}
\label{subsec:housing}
As previously discussed, neutrino telescopes are typically located deep in water or ice to provide not only a Cherenkov medium but also shielding against atmospheric background. Thus, a potential calibration device has to be able to withstand high pressure (up to 250 atm) and cold temperatures (down to -40C). In partnership with the German marine housing specialist Nautilus Marine Service GmbH, we developed a titanium pressure housing providing close to $4\pi$ light emission field-of-view (FOV), a theoretical pressure resistance up to around $1,000\,$atm, a temperature resistance down to $-55^\circ\,$C and a vibration- and shock resistance according to ISO 13628-6. The cylindrical design shown in \cref{fig:housing-all} allows electronics and necessary cable penetrators for supply and communication to be placed outside of its light emission FOV. Especially important was the design of the two flanges as they pose the major influence on the eventual emission profile of the instrument. This will be discussed in detail in \cref{subsec:emissionprofile}. In general, the flanges are used to complement the pressure housing with optically-enhanced borosilicate glass~\cite{nbk7}, attached via deep-sea epoxy resin~\cite{epoxy}. Furthermore, they also provide means of mounting electronics internally to each of the instrument sides. The instrument measures around $40\,$cm in length and $13\,$cm in diameter, excluding penetators.

We have applied several ‘stress tests’ to verify the performance of the POCAM under environmental conditions. For pressure testing, the sealed housing was cycled repeatedly in a pressure chamber to $700\,$bar (690$\,$atm) at a facility provided by Nautilus. For temperature testing, it was placed in a freezer at $-55\,^\circ$C for a total period of five weeks. In both cases internal pressure, temperature and humidity were monitored throughout the testing period and were found to remain stable. Vibration and shock tests were performed and passed at the German test facility IABG.

\subsection{Light pulsers}
\label{subsec:pulsers} 
\begin{figure}[h!]
	\centering
	\begin{tikzpicture}
	\draw[draw=white, fill=gray!20] (-7.5,-2.5) rectangle (7.5, 2.5);
	\draw[] (0.5,1) -- (3,1);
	\draw[] (1.75,0.5) -- (3,1);
	\draw[] (1.75,-0.5) -- (3,1);
	\draw[] (0.5,-1) -- (3,1);
	\draw[dotted, thick] (0.5,-1) -- (3,-1);
	\draw[dotted, thick] (1.75,0.5) -- (3,-1);
	\draw[dotted, thick] (1.75,-0.5) -- (3,-1);
	\draw[dotted, thick] (0.5,1) -- (3,-1);
	\draw[] (-1.75,0.5) -- (-3,1);
	\draw[] (-1.75,-0.5) -- (-3,1);
	\draw[dotted, thick] (-1.75,-0.5) -- (-3,-1);
	\draw[dotted, thick] (-1.75,0.5) -- (-3,-1);
	\draw[draw=black, fill=red!20] (-0.5,0.5) rectangle (0.5,1.5);
	\draw[draw=black, fill=green!20] (-1.75,0.125) rectangle (-0.75,1.125);
	\draw[draw=black, fill=red!20] (0.75,0.125) rectangle (1.75,1.125);
	\node[] at (-1.25,0.625) {$\lambda_1$};
	\node[] at (0,1) {$\lambda_1$};
	\node[] at (1.25,0.625) {$\lambda_2$};
	\draw[draw=black, fill=green!20] (-0.5,-0.5) rectangle (0.5,-1.5);
	\draw[draw=black, fill=green!20] (-1.75,-0.125) rectangle (-0.75,-1.125);
	\draw[draw=black, fill=red!20] (0.75,-0.125) rectangle (1.75,-1.125);
	\node[] at (-1.25,-0.625) {$\lambda_5$};
	\node[] at (0,-1) {$\lambda_4$};
	\node[] at (1.25,-0.625) {$\lambda_3$};
	\fill[green!20] (-1.55,-1.9) rectangle (-0.75,-2.2);
	\node[] at (-0.35, -2.05) {\scriptsize LED};
	\fill[red!20] (0.25,-1.9) rectangle (1.05,-2.2);
	\node[] at (1.35, -2.05) {\scriptsize LD};
	\draw[draw=black, fill=white!10] (-6.75,0.25) rectangle (-3,1.75);
	\draw[draw=black, fill=white!10] (-6.75,-0.25) rectangle (-3,-1.75);
	\node[] at (-4.875,1) {LED driver $\#$1};
	\node[] at (-4.875,-1) {LED driver $\#$2};
	\draw[draw=black, fill=white!10] (6.75,0.25) rectangle (3,1.75);
	\draw[draw=black, fill=white!10] (6.75,-0.25) rectangle (3,-1.75);
	\node[] at (4.875,1) {LD driver $\#$1};
	\node[] at (4.875,-1) {LD driver $\#$2};
	\end{tikzpicture}
	\caption[POCAM LED layout]{POCAM LED and pulse driver layout. Four diodes are driven with an LD-type driver, two LEDs with an optimized Kapustinsky driver. Both driver types exist on the board twice for redundancy and can be selectively enabled for a specific emitter wavelength $\lambda_i$. For details refer to the text.}
	\label{fig:led-layout}
\end{figure}
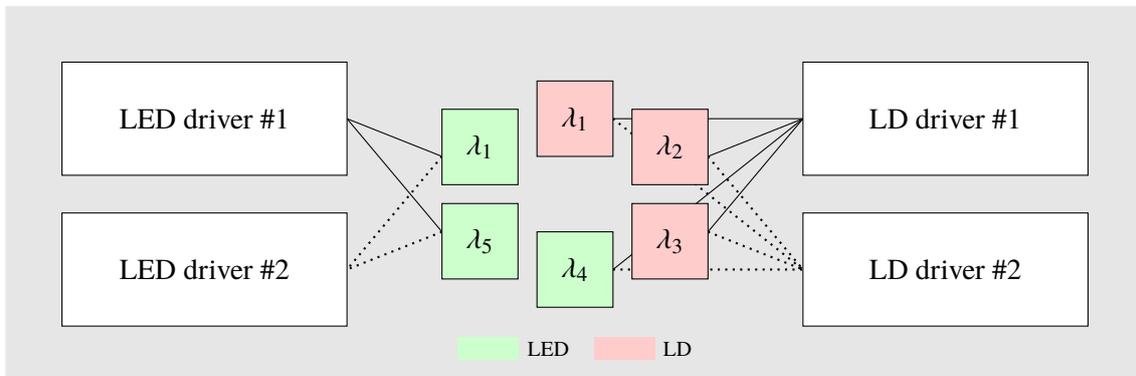

In total the POCAM can host six light emitters which in turn can be controlled by two different light pulsers each. This concept is visualized in \cref{fig:led-layout}. The typical choice of optical emitters are light emitting diodes (LEDs) as well as laser diodes (LDs). Due to the support for two pulse driver types, the POCAM generally makes use of two distinct pulse drivers: the Kapustinsky circuit~\cite{KAPUSTINSKY1985612} and a LD-type LD driver~\cite{epc9126hc}. By using switching components and circuitry, we allow the usage of a specific emitter with a different number of pulse drivers. This not only allows for redundancy in the field but also enables us to include various emitter wavelengths as they can then be selectively driven by different pulsers.
\par
\begin{figure}[ht!]
	\centering
	\includegraphics[]{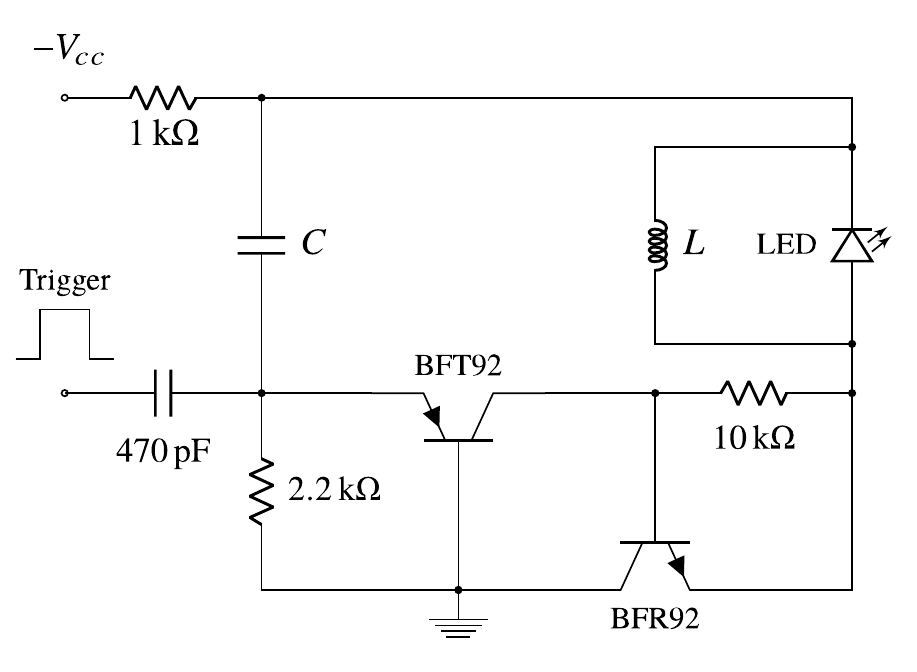}
	\caption[Kapustinsky flasher schematic.]{Kapustinsky flasher schematic for the POCAM. The circuit is operated on negative bias voltage and produces a pulse when triggered with a square pulse signal. The light pulse is mainly shaped by the capacitor $C$ and the inductance $L$ as well as the LED itself.}
	\label{fig:kapustinsky-schematic}
\end{figure}

The Kapustinsky circuit is optimized for LEDs and makes use of a controlled capacitor discharge which is artificially cut short by an inductance parallel to the emitter. This pulser has the great advantage of linear light output adjustable with bias voltage and a relatively simple circuitry which is shown in \cref{fig:kapustinsky-schematic}. Pulse properties can be pre-selected by adjusting the values of the discharged capacitor $C$ and the used inductance $L$, and the pulse width scales qualitatively with $\sqrt{LC}$. In addition to being easy to set up, it is further used in related experiments~\cite{Bedard:2018zml,nanobeacon} and so comes with the benefit of experience and reliability. In the current configuration, the Kapustinsky circuit drives $405\,$nm and $465\,$nm LEDs per optimization for its IceCube application. For the POCAM application in the IceCube Upgrade we will host two distinct and switchable Kapustinsky drivers per LED with respective $(L,\,C)$ values of $(22\,\text{nH}, 100\,\text{pF})$ for the fast and $(22\,\text{nH}, 1.2\,\text{nF})$ for the default pulse configuration.
\par
\begin{figure}[ht!]
    \centering
	\includegraphics[]{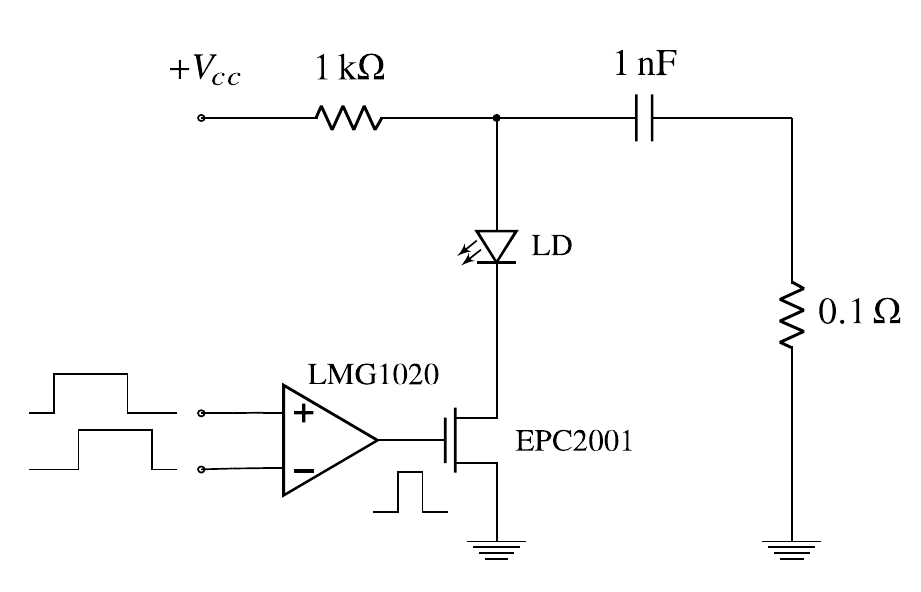}
	\caption[LD driver flasher schematic.]{LD driver flasher schematic for the POCAM. The circuit is operated on positive bias voltage and produces a pulse proportional to the input signal. The light pulse is mainly shaped by the input pulse and the bias voltage as well as the LED itself.}
	\label{fig:lidar-schematic}
\end{figure}

\noindent
The LD-type driver is an industry standard frequently used for LIDAR applications~\cite{epc9126hc} and makes use of high-current ultra-fast switching of Gallium-Nitride field effect transistors (GaN-FETs). The GaN-FETs allow picosecond switching of voltages up to $100\,$V and currents up to $40\,$A which are typically supplied by a bank of capacitors and discharged through the LD. This typically results in very short and intense light pulses if used with appropriate emitters. The general circuitry concept is shown in \cref{fig:lidar-schematic}. As for emitters, for this circuit we are limited to available LD wavelengths with sufficient performance. However, typical wavelengths between $400\,$nm and $600\,$nm with matching characteristics are readily available. It is also useful to drive UV LEDs with these circuits as they typically require much larger currents to produce significant light outputs. In the current configuration the emitters include an LED at $365\,$nm as well as LDs at $405\,$nm, $455\,$nm and $520\,$nm. A summary of selected emitters is given in \cref{tab:emitters} together with available drivers.

\begin{figure}[h!]
	\centering
	\hspace*{-1cm}
    \includegraphics[width=0.9\textwidth]{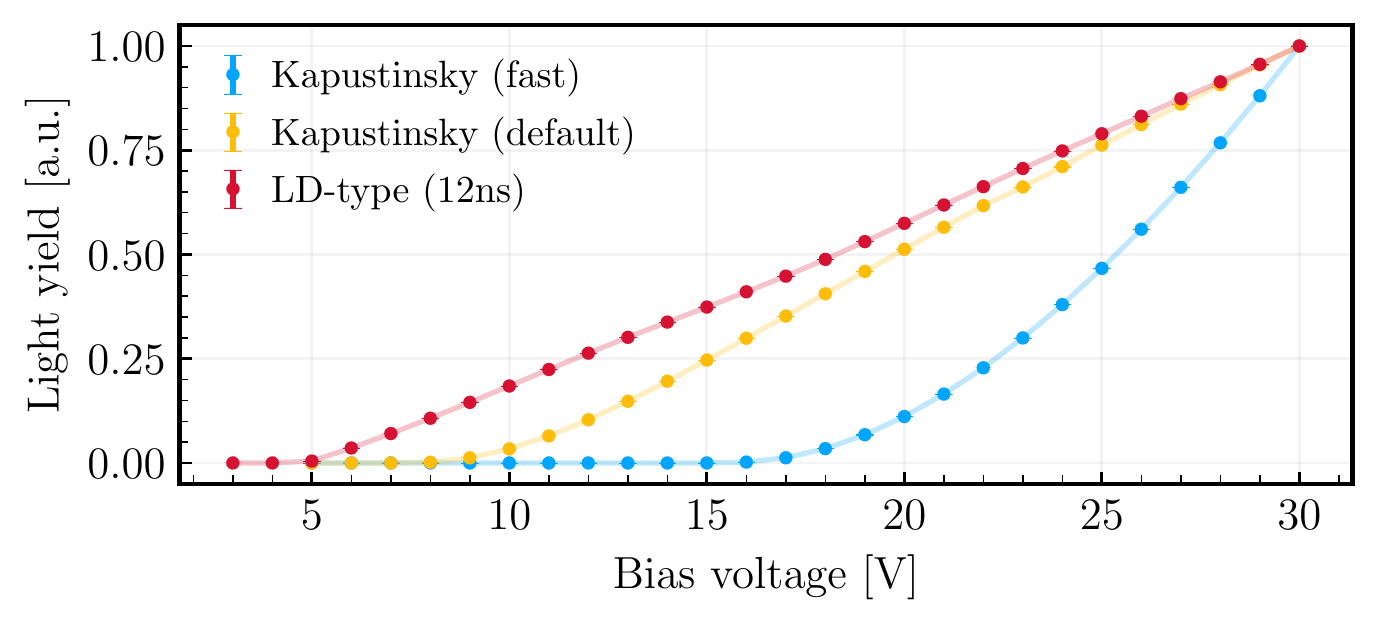}
    \caption{POCAM light pulser linearity. The figure shows the normalized intensity of all light pulse drivers with respect to applied bias voltage for the default $405\,$nm LED/LD at room temperature.}
    \label{fig:pulser-linearity}
\end{figure}

\begin{figure}[h!]
    \centering
    \hspace*{-1cm}
    \includegraphics[width=0.9\textwidth]{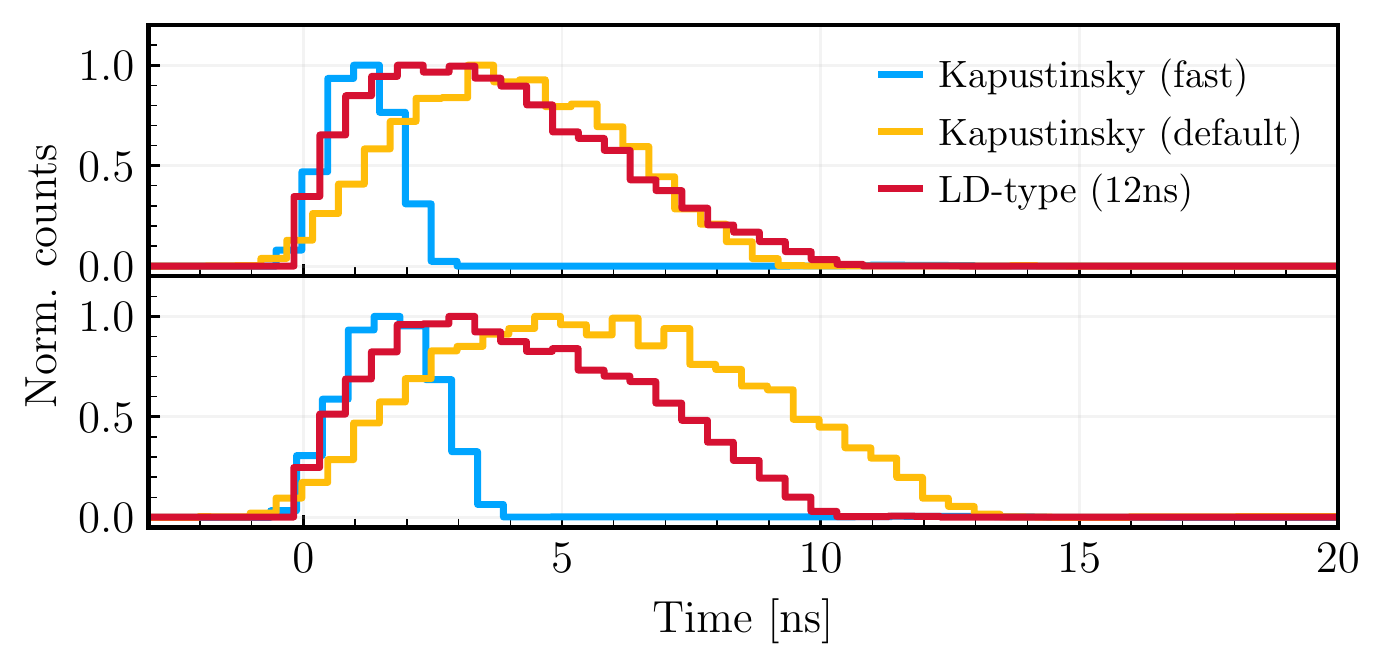}
    \caption{POCAM light pulser time profiles. The figure shows the normalized time profiles of all light pulse drivers with respect to minimal-working (top) and maximal (bottom) applied bias voltage for the default $405\,$nm LED/LD at room temperature.}
    \label{fig:pulser-timing}
\end{figure}

\begin{table}[ht!]
	\centering
	\begin{tabular*}{\textwidth}{l @{\extracolsep{\fill}} c c c}
		\toprule 
		\bf Emitter & \bf Wavelength [nm] & \bf Kapustinsky & \bf LD-type\\
		\midrule
		XSL-365-5E~\cite{365nm} & $365$ & -- & 2x \\
		XRL-400-5E~\cite{400nm} & $405$ & fast / default & --\\
		RLT405500MG~\cite{ld405nm} & $405$ & -- & 2x \\
		PL-TB450B~\cite{ld450nm} & $450$ & -- & 2x \\
		NSPB300B~\cite{470nm} & $465$ & fast / default & --\\
		LD-520-50MG~\cite{ld520nm} & $520$ & -- & 2x \\
		\bottomrule
	\end{tabular*} 
	\caption[]{POCAM emitter selection with corresponding wavelengths and available drivers. Each LED driven by a Kapustinsky can select a fast or default pulser configuration with different pulse widths and light yields. Each LD-type driven diode can select one of the two redundant LD-type drivers.}
	\label{tab:emitters}
\end{table}
Both circuits perform in different regions of the dynamic range of the POCAM. With the fast Kapustinsky driver being the dimmest followed by the default Kapustinsky driver, the LD driver generally outperforms in terms of intensity but at the marginal cost of longer pulse widths. The normalized intensity behavior of all circuits and an emitter wavelength of $405\,$nm is shown in \cref{fig:pulser-linearity}, and their corresponding timing performance in \cref{fig:pulser-timing}. While the former has been measured with an external photodiode, the latter was done using time-correlated single photon counting with an avalanche photodiode (APD) to sample the pulse time profile. It is evident that the intensity behavior is linear for all circuits over the major parts of their dynamic range with estimated total number of photons for each pulser at maximum brightness given in \cref{tab:pulsers}. Furthermore, as seen in \cref{fig:pulser-timing}, the timing behavior of the Kapustinsky circuits has been fine-tuned by selection of respective ($L, C$) values to FWHMs of around $1-3\,$ns and $4-8\,$ns, for the fast and default configuration, respectively. The LD driver on the other hand, can generate a wide range of pulse lengths by means of enable and disable inputs together with a programmable delay line. The only limitation here is to avoid current or heat damage of the emitter. The operational range of pulse settings was set to be adjusted between $1.5 - 35\,$ns in FWHM input pulse widths for the LDs. A summary of typical FWHMs achieved with different pulsers is also given in \cref{tab:pulsers}. It should be noted that the circuit performance is heavily influenced by the selected emitter and it is generally necessary to perform appropriate selection procedures to identify suitable choices. This can even be dependent on production year with varying semiconductor batches for LEDs and LDs. These emitters further give need for appropriate calibration procedures, especially with respect to ambient temperature. The experimental setups and prototype results from calibrations are explained in \cref{subsec:pocamcalibration}. Lastly, both circuits support flashing frequencies up to at least a few tens of kHz but were optimized for a maximum flashing frequency of $1\,$kHz for the IceCube Upgrade application.
\par
\par\noindent
\begin{table}[h!]
\centering
\begin{tabular*}{\textwidth}{l @{\extracolsep{\fill}} c c c c c}
\toprule
\textbf{Pulser} & \textbf{Emitter [nm]} & \multicolumn{2}{c}{\textbf{Low intensity}} & \multicolumn{2}{c}{\textbf{High intensity}} \\
 & \multicolumn{1}{l}{} & \multicolumn{1}{l}{Photons} & \multicolumn{1}{l}{FWHM {[}ns{]}} & \multicolumn{1}{l}{Photons} & \multicolumn{1}{l}{FWHM {[}ns{]}} \\ \midrule
\multirow{2}{*}{\textbf{Kapustinsky (fast)}} & 405 & $ 6.3 \times 10^7$ & 1.4 & $ 2.7 \times 10^8$ & 2.4 \\
 & 465 & $1.2 \times 10^7$ & 2.6 & $5.7 \times 10^7$ & 3.1 \\ \midrule
\multirow{2}{*}{\textbf{Kapustinsky (default)}} & 405 & $ 8.5 \times 10^8$ & 4.9 & $ 4.0 \times 10^9$ & 8.0 \\
 & 465 & $6.0 \times 10^8$ & 7.6 & $2.5 \times 10^9$ & 10.5 \\ \midrule
\multirow{4}{*}{\textbf{LD-type (5ns)}} & 365 & $1.3 \times 10^9$ & 6.3 & $1.9 \times 10^9$ & 6.2 \\
 & 405 & $5.9 \times 10^9$ & 3.7 & $2.9 \times 10^{10}$ & 4.0 \\
 & 450 & $1.1 \times 10^{10}$ & 4.4 & $4.7\times 10^{10}$ & 3.8 \\
 & 520 & $3.2 \times 10^9$ & 11.1 & $1.4 \times 10^{10}$ & 6.2 \\ \midrule
\multirow{4}{*}{\textbf{LD-type (25ns)}} & 365 & $3.0 \times 10^9$ & 7.3 & $7.8\times 10^9$ & 10.0 \\
 & 405 & $4.6 \times 10^{10}$ & 5.5 & $1.3 \times 10^{11}$ & 5.9 \\
 & 450 & $2.1 \times 10^{10}$ & 5.4 & $1.0 \times 10^{11}$ & 7.3 \\
 & 520 & $5.5 \times 10^9$ & 9.7 & $2.7 \times 10^{10}$ & 12.4 \\
 \bottomrule
\end{tabular*}%
\caption{Exemplary bare light pulser performance for all wavelengths at low and high intensity settings at room temperature. The LD driver is further shown for two input pulse widths as it scales the brightness but also time profile of the emitted pulses. The photon number was measured using a photodiode and represents a lower limit on the emission with a systematic uncertainty of approximately $10\,\%$.}
\label{tab:pulsers}
\end{table}\par\noindent

\subsection{Isotropy}
\label{subsec:emissionprofile}
One primary aspect of the POCAM is its isotropical emission profile. This not only guarantees intrinsic independence of its orientation but further allows parallel self-monitoring of its light output. Thus the POCAM is independent of the observing detector, and hence suffers reduced loop-back uncertainties resulting from imperfect detector understanding. The general concept here makes use of the aforementioned light pulsers for generation of light pulses which are then diffused by an integrating sphere made from teflon. Here, a two-part geometry ensures that the light pulse is pre-diffused by a teflon plug before being integrated in the remainder of the sphere, ensuring a higher grade of isotropy. The latter is further made from optical teflon~\cite{berghof} which has been shown to drastically improve the isotropy. And, as will be discussed in \cref{subsec:selfmonitoring}, self-monitoring is then done by integrated photosensors which are provided fixed solid angles by the aperture mounting. The latter comprise of two half-disks and a mounting ring, all from stainless steel, to mount the diffusing spheres and provide fixed solid angles for the sensors. Stainless steel is the material of choice as it provides a similar thermal expansion coefficient to the titanium in the pressure housing. 

\begin{figure}[h!]
	\centering
	\includegraphics[width=1\textwidth]{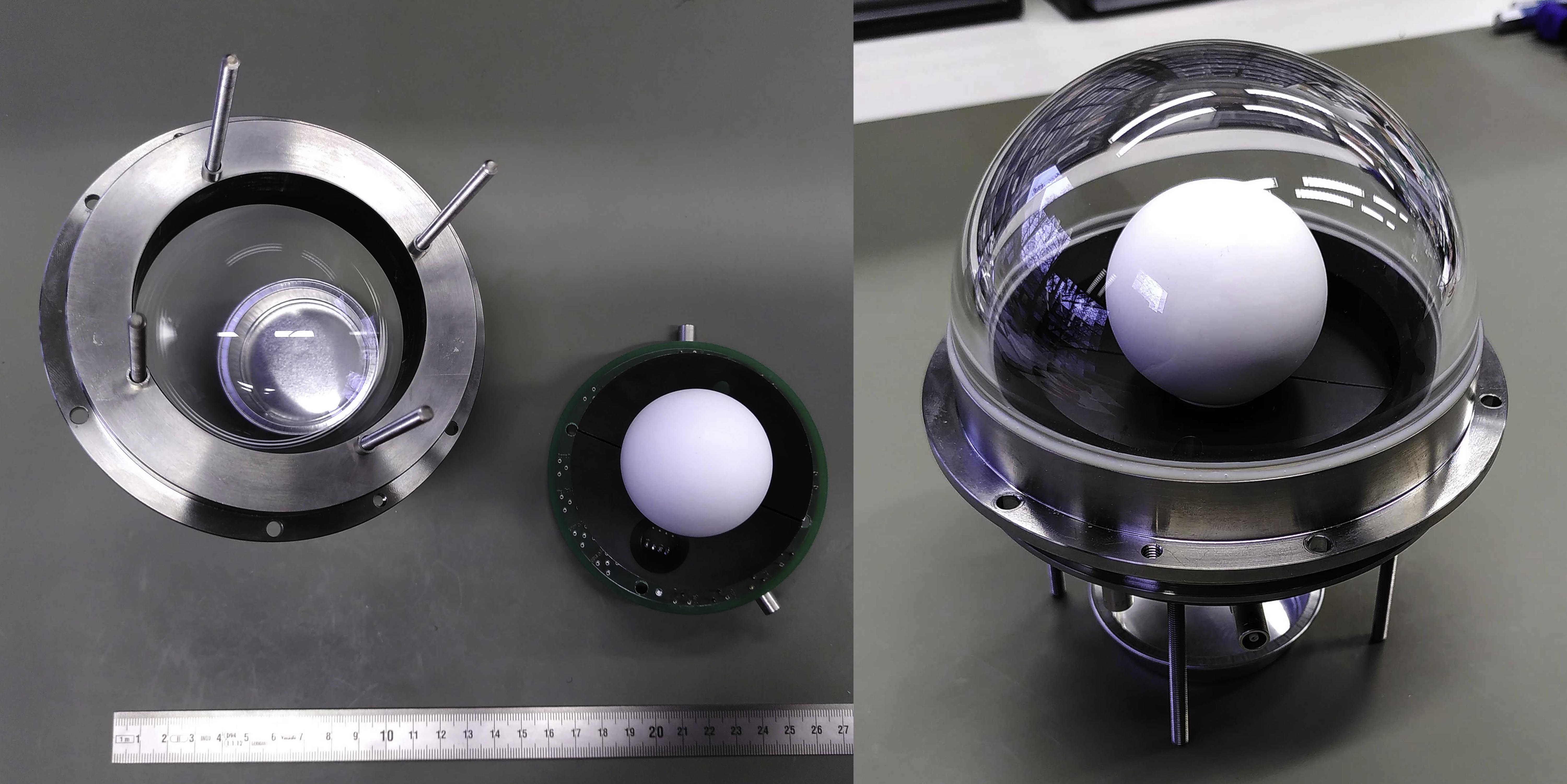}
	\caption{POCAM hemisphere prototype during assembly (left) and assembled (right). For details on the components of the visible prototype analog board refer to \cref{fig:housing-flange} and the text.}
	\label{fig:hemisphere-real}
\end{figure}

\begin{figure}[h!]
	\centering
	\hspace*{-1cm}
	\includegraphics[width=0.9\textwidth]{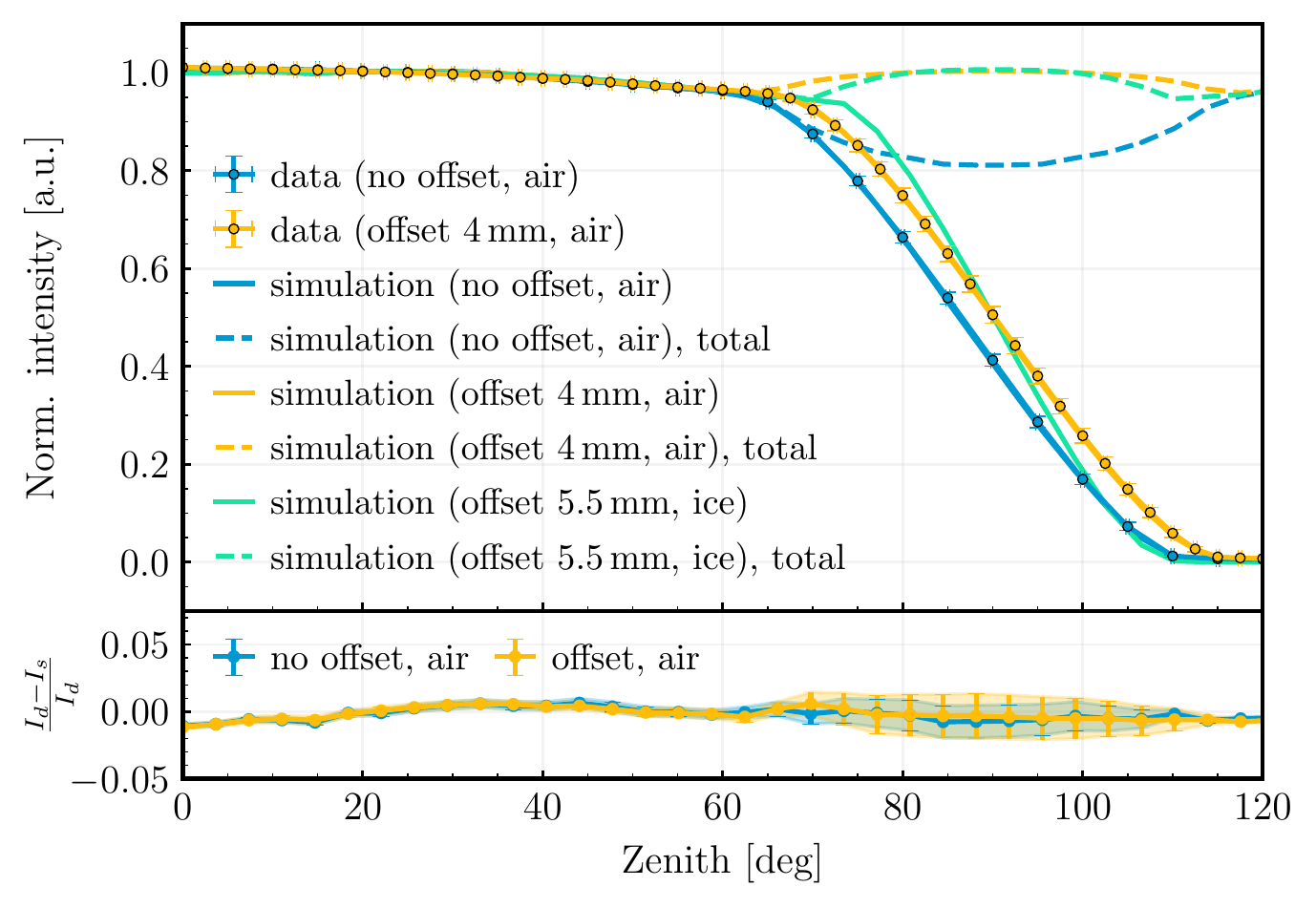}
	\caption[]{Measured and simulated emission profile of a POCAM hemisphere assembly. The top part of the figure shows the measured and simulated intensity per solid angle, given as a function of zenith angle with respect to the POCAM cylinder axis. Both simulations and measurements are shown for the single hemisphere alone (solid) and for a virtual sum of two mirrored hemispheres at infinite distance (dashed) including respective data. The bottom plot shows the normalized difference between data and simulation with maximum deviations of around $1\,$\%. The data errorbars represent the $1\sigma$ errors from azimuthal deviations.}
	\label{fig:emission-geant} 
\end{figure}

It should be noted that low-reflective coating of all internal components is required in order to avoid adding a cosine-like reflective component to the emission profile which would skew the achieved isotropy. This is realized using MagicBlack coating~\cite{magicblack} which is applied to both the apperture disks as well as the internal flange surfaces. A hemisphere assembly is shown in \cref{fig:hemisphere-real} together with a view during assembly of sub-components.
The flange design poses a significant influence on the emission profile and hence its design was critical. Especially since the glass hemispheres are attached to the flange via epoxy, we required the manufacturer to provide well controlled and small edges of the latter around the waistband. This was shown necessary by measurements using a dedicated calibration setup in order to avoid significant deviations from isotropy around the waistband as well as azimuthal deviations. The setup itself is explained in detail in \cref{subsec:pocamcalibration}. To investigate optimizations of the system, a GEANT4~\cite{AGOSTINELLI2003250} simulation framework of the POCAM was setup which was able to reproduce the measured emission profile to within $1\,\%$. This is shown in \cref{fig:emission-geant} together with measured emission profile data of a single hemisphere assembly. The simulation further showed that a small remaining glue edge could be counteracted by offsetting the integrating sphere upwards with respect to the equator of the hemisphere assembly, which was further confirmed by measurements. However, this simulation framework is also necessary to relate the calibration measurements done in air to the actual emitted light profile into water or ice. Measurements of the emission profile in water as well as a complete $4\pi$ scan are currently underway and are expected to confirm the simulated results. 
\par\noindent

\subsection{Self-monitoring}
\label{subsec:selfmonitoring}
\begin{figure}[h!]
    \centering
    \begin{subfigure}[b]{0.45\textwidth}    	
        \centering
    	\scalebox{1}{%
	    \includegraphics[]{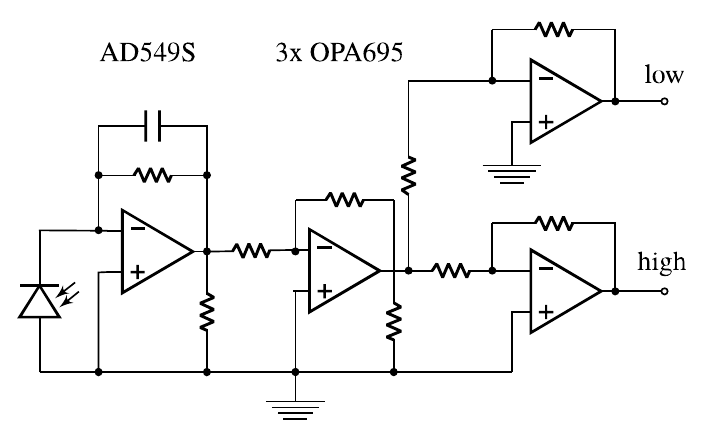}
	    }
    	\caption[]{Transimpedance charge readout for the photodiode with two gain stages.}
    	\label{fig:insitu-readouts-pd}
    \end{subfigure}\hspace{0.025\textwidth}
    \begin{subfigure}[b]{0.5\textwidth}
    	\centering
    	\scalebox{1}{%
	    \includegraphics[]{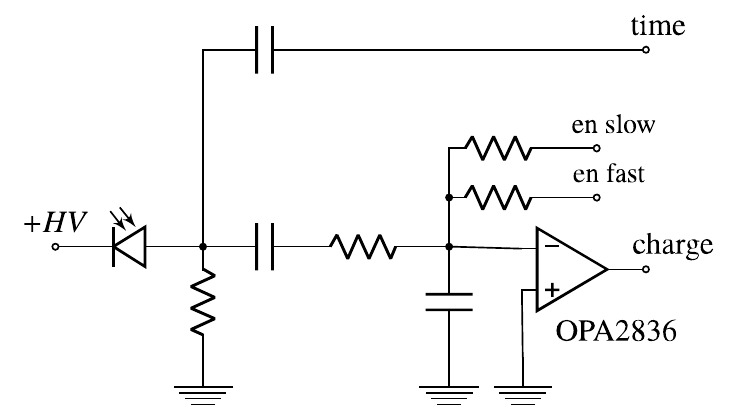}
	    }
    	\caption[]{GSI-developed charge and time readout of the SiPM fed into an FPGA-based discriminator.}
    	\label{fig:insitu-readouts-sipm}
    \end{subfigure}
    \caption{Conceptual views of the self-monitoring sensor readouts for the PD and the SiPM used in the POCAM. For details on their functionality refer to the text.}
    \label{fig:in-situ-readouts}
\end{figure}

In order to provide independent monitoring from the telescope, the POCAM includes self-monitoring photosensors. These sensors monitor its per-pulse light output and give a handle for correction of intensity fluctuations over the course of its deployment period. The photosensors are two-fold: we use a Silicon-photomultiplier (SiPM) for low light intensity and a photodiode (PD) for high light intensity. The chosen SiPM is the \emph{Ketek 3315-WB}~\cite{sipm} and the PD is the \emph{Hamamatsu S2281-01}~\cite{pd}. As discussed in \cref{subsec:emissionprofile}, the internal mounting of the diffusing spheres is done using aperture disks from stainless steel. These disks provide the solid angles for the photosensors and further means of mounting for the PD. The SiPM measures low charges as well as the pulse on-set time and its approximate duration at $350\,$ps binning, the PD measures integrated charge over a large dynamic range, pre-dominantly aimed at higher intensities.~Both sensors use dedicated charge readout schemes to provide intensity information with relevant parts of the circuit are shown in \cref{fig:in-situ-readouts}.
\par
\begin{figure}[h!]
	\centering
	\begin{subfigure}[b]{0.9\textwidth}
	    \centering
	    \hspace*{-1cm}
	    \includegraphics[width=\textwidth]{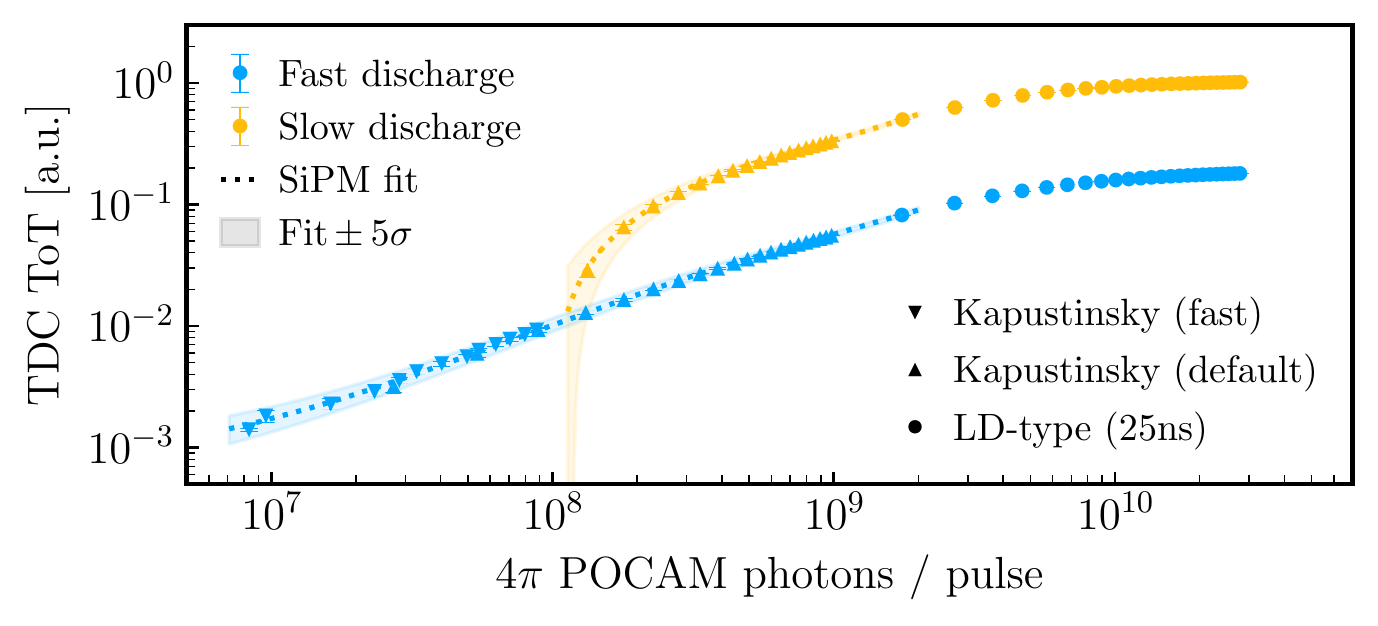}
	    \caption{SiPM readout responses and fits for both fast and slow channels of the TDC circuit in an assembled POCAM hemipshere using all pulser types at $405\,$nm and a SiPM over-voltage of $5\,$V.}
	\end{subfigure}
	\par\vspace{6pt}
	\begin{subfigure}[b]{0.9\textwidth}
	    \centering
	    \hspace*{-1cm}
	    \includegraphics[width=\textwidth]{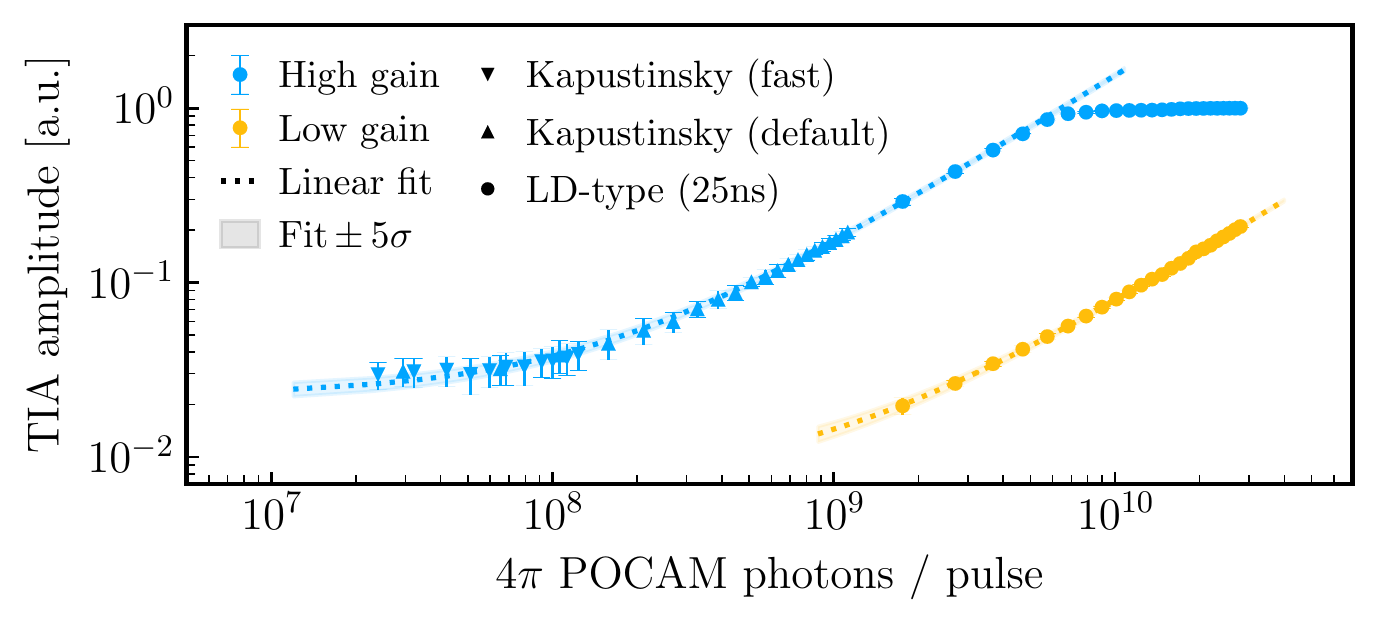}
	    \caption{Normalized photodiode readout responses and linear fits for both gain channels of the TIA circuit in an assembled POCAM hemipshere and using all pulser types at $405\,$nm. The high-gain channel shows saturation at highest intensities but is compensated by the low-gain channel.}
	\end{subfigure}
	\caption{Self-monitoring sensor response for both the SiPM (top) as well as the photodiode (bottom) as a function of total POCAM light emission including respective fits for all pulser types at $405\,$nm and room temperature. The errorbars represent the $1\sigma$ pulse-to-pulse spread of the measured sensor responses and the given photon number is subject to a systematic uncertainty of approximately $10\%$ in the used setup.}
	\label{fig:insitu-readouts-linearity}
\end{figure}

\noindent
 For the SiPM, we make use of a time-to-digital-converter (TDC) circuitry developed together with the GSI Helmholtz Centre for Heavy Ion Research GmbH in Darmstadt, Germany and the TRB-Collaboration (trb.gsi.de). This circuit enables the measurement of both on-set time and charge information by correlated pulse splitting and controlled discharge of a capacitor. Both signals are then fed into an FPGA-based discriminator and eventually the TDC for digitization with charge proportional to the signal time-over-threshold (ToT). For the PD we make use of a traditional transimpedance amplifier (TIA) circuit using an extremely low-noise amplifier AD549S~\cite{ad549s} which provides a voltage amplitude proportional to the measured charge of the PD. The TIA is then followed by two additional low- and high-gain amplifier stages which are fed into an analog-to-digital converter (ADC). These are necessary in order adjust to the DAQ input voltage range limits and to tune the signals and their offsets appropriately. The goal of these circuits is to eventually provide a dynamic range of self-monitoring which is able to cover the full range of emission intensities with good precision and their performance has been optimized in that respect, with response behavior shown in \cref{fig:insitu-readouts-linearity}. While the PD was expected to be linear across all its measurable range~\cite{pmid28184106}, the SiPM shows the expected saturation behavior but can easily be fit with appropriate exponential functions~\cite{Kotera:2015rha,Rosado:2017ebu}. Aging of the sensors, while not expected for either the  SiPM~\cite{Sun:2015pwr,2008JInst...310001L} nor the PD due to low light levels~\cite[e.g.][]{Eppeldauer:90}, will be tested in long-term test stands.

\subsection{DAQ, control and power electronics}
\label{subsec:electronics}
The POCAM electronics generally consist of a mirrored system for both hemispheres with an analog and digital front-end board. Those contain the components necessary for light pulsing, self-monitoring as well as their power supplies and control by means of an FPGA. In addition to these mirrored systems, this POCAM iteration hosts three IceCube-specific boards. The wire pair originating from the IceCube network is first handled by the \textit{IceCube communications module} (ICM)~\cite{Nagai:2019uaz} which interfaces the system to the IceCube data-, clock- and power stream and takes care of providing the proper communication protocols, synchronization signals and power. The ICM signal chain is then fed to a secondary board which provides both hardware and software similar to the DAQ of IceCube Upgrade sensor instruments with a micro-controller unit (MCU) in its core. A third distribution board distributes data, clock, and communication streams as well as power to each hemisphere using an FPGA. The first two interface boards are aimed specifically for an application in the IceCube Upgrade but can be exchanged to fit any telescope-specific back end. In this configuration the POCAM was designed to be powered with $96\,$V and a maximum power consumption of $8\,$W.  As for functionality, the FPGA in each hemisphere receives slow-control commands and clocks from the MCU via the distribution board. The slow control prepares the FPGA registers according to the measurement run to be carried out and sets up the configuration parameters for the flashers as well as the self-monitoring sensors and their DAQ. Once the flash is initiated, the self-monitoring sensors monitor the per-pulse intensity as well as its timing behavior and write this data to the FPGA internal storage. After the flashing procedure has finished, the MCU can access this data and eventually transfer it via the IceCube data stream for offline processing. While the major part of the POCAM-specific digital electronics were already successfully tested and optimized in GVD and STRAW, the back end interfaces are a new addition for the IceCube Upgrade and are currently in advanced prototyping. 

\subsection{Calibration}
\label{subsec:pocamcalibration}
In order to streamline the calibration process for the production phase of around 30 POCAMs, we developed two dedicated setups to characterize the flashers and the emission profile of each hemisphere, respectively. The former is a relative flasher characterization setup with the goal to provide knowledge of the flasher's relative intensity, time profile and spectrum variation as a function of configuration parameters and temperature. The latter is an emission profile characterization setup which aims to characterize the emitted light pattern of each POCAM hemisphere relatively. 

\begin{figure}[h!]
	\centering
	\includegraphics[width=\textwidth,trim=1.6cm 1.5cm 0.8cm 3cm,clip]{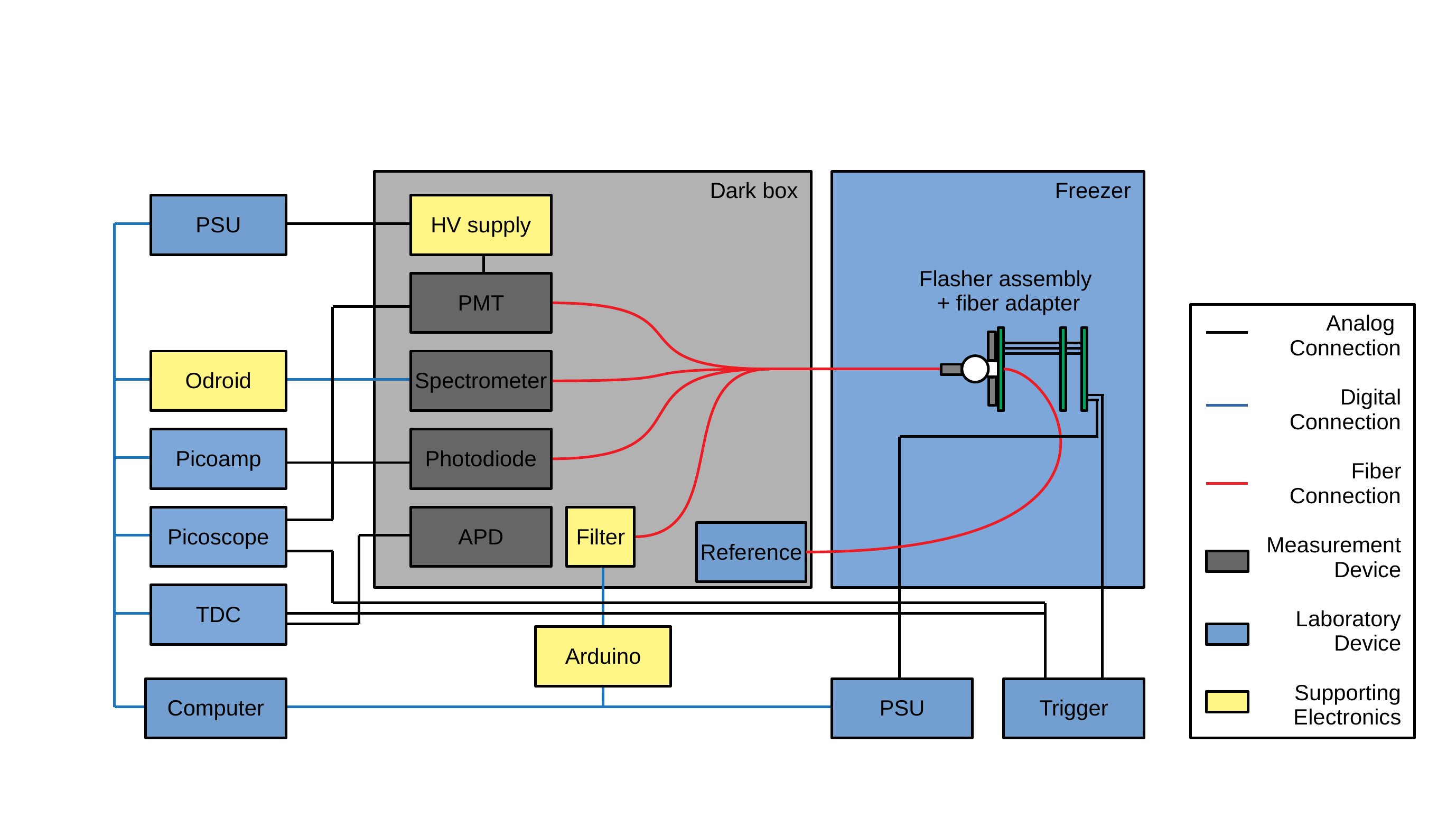}
	\caption{Schematical workflow diagram of the light pulser calibration station. For details on the sub-components and their functions, refer to the text.}
	\label{fig:pocas-schematic}
\end{figure}

\begin{figure}[h!]
	\centering
	\includegraphics[width=0.985\textwidth,trim=2.5cm 5.2cm 1cm 2.5cm,clip]{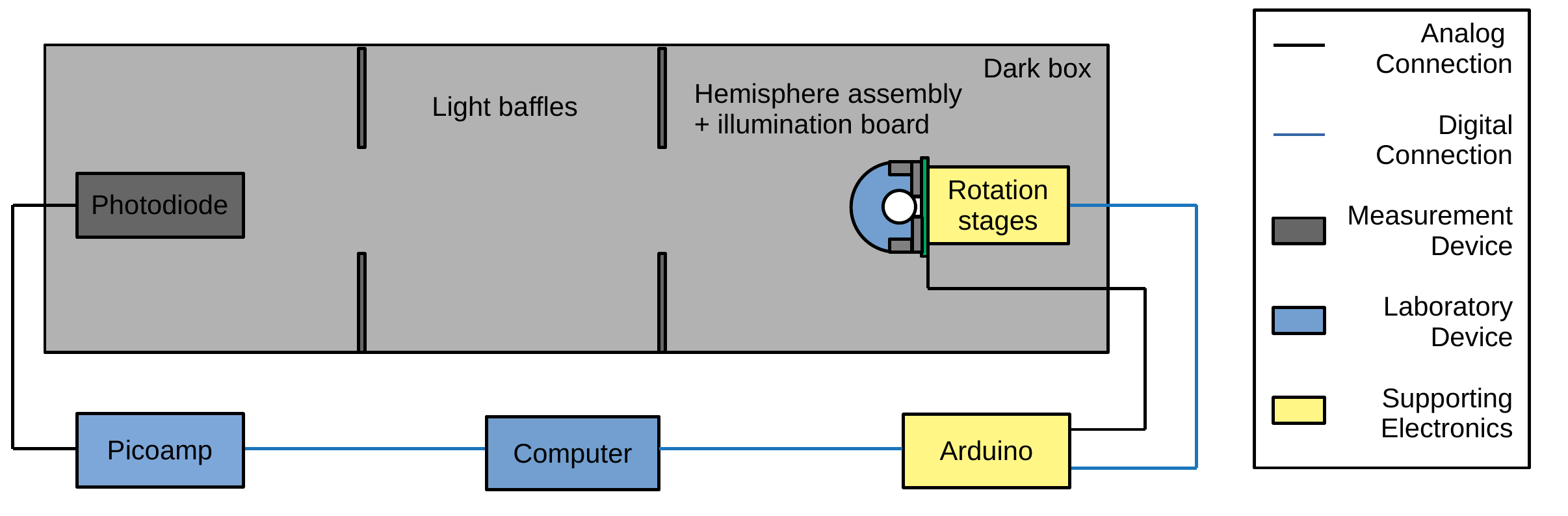}
	\caption{Schematical workflow diagram of the emission profile calibration station. For details on the sub-components and their functions, refer to the text.}
	\label{fig:angcas-schematic}
\end{figure}

\paragraph*{Relative flasher characterization} The flasher characterization setup, shown in \cref{fig:pocas-schematic}, consists of four sensors: a photodiode~\cite{pd}, a PMT~\cite{pmt}, an avalanche photodiode (APD)~\cite{apd} and a spectrometer~\cite{spec}. These sensors are located in a dark box and each is coupled to one end of a 4-to-1 fan-out fiber, the single end of which is coupled to a flasher assembly including diffusers and apperture disks. This assembly itself is further located in a freezer together with the remainder of the POCAM electronics and can be cooled down to $-75\,^\circ$C. While the PD and PMT record the intensity, the APD uses time-correlated single photon counting to measure the pulse time profile and the spectrometer directly records the output spectrum. The APD further uses a controllable neutral density filter wheel to achieve low occupancies of below $10\,\%$ which is necessary to provide proper single-photon sampling of the time profile. The raw PMT pulses are recorded with the help of a digital oscilloscope, the PD is read out with the \emph{Keithley 6485} picoammeter, the APD is fed into a high-precision TDC and the spectrometer outputs the spectrum via serial command. Together with necessary power supplies and other peripheral electronics, all of the sub-components are controlled by a dedicated computer running all the necessary software. The procedure starts by mounting a specific POCAM inside the freezer and, using a dedicated stainless steel structure, coupling the fan-out fiber to the diffuser. Then, the automated system scans temperature and configuration parameters and measures the pulse properties as well as the self-monitoring sensor responses. This data can then be used offline to provide individual relative characterization of a specific POCAM hemisphere assembly and results in fingerprint-characterized or \emph{golden} POCAMs. To further remove systematic uncertainties resulting from potential fiber coupling changes over the course of cooling and heating, we flush the freezer with nitrogen as well as monitor a temperature-stabilized reference halogen light source coupled into the diffuser using the same type of fiber.
\paragraph*{Relative emission profile characterization} The emission profile setup consists of a two-axis rotation stage assembly which allows mounting a POCAM hemisphere including flange, diffusers, apperture disk and sensors in a dark box of around $140\,$cm inner length. The rotation stages used provide sub-degree precision and two of them are used to create a custom two-axis rotation stage. On the opposite side, a photodiode~\cite{pd} is mounted and light baffles in between further reduce stray light from reflections off of inner surfaces. The hemisphere is then mounted to the rotation stages with a dedicated illumination board which provides identical layout to the analog board but uses the LEDs in switchable forward mode. A dedicated measurement PC then controls the characterization scan for a set of azimuth and zenith angles as well as LEDs and measures the intensity data of the PD. The PD current is monitored at each angle step with a \emph{Keithley 6485} picoammeter for both every LED switched on individually and every LED switched off. This data is then eventually written to file. Due to the rotation of the hemisphere, this provides a relative characterization of its emission profile and can further be used to calculate the total hemispherical light yield. The workflow of this setup is shown in \cref{fig:angcas-schematic} including peripheral electronics.
\paragraph*{Absolute calibration} After all POCAMs have undergone relative characterization, the last step is absolute calibration of the light yield. For this, the instruments are placed in the emission profile setup and the regular PD is exchanged for one precision-calibrated by the National Institute for Standards and Technology (NIST). Then, iterating through all emitters and pulse drivers as well as a range of configuration parameters and measuring the photocurrent at the PD, this provides absolute intensity scales to all previous relative calibrations. These results are then the reference for self-monitoring data over the course of the operational period of the instrument and can be used for correcting the instrument emission in-situ using the integrated photosensors. The intrinsic systematic uncertainties of this calibration chain are given in \cref{tab:systematics} including the total anticipated systematic uncertainty on the POCAM light yield of $4.1\,\%$.
\begin{table}[h!]
\centering
\begin{tabular*}{\textwidth}{l @{\extracolsep{\fill}} l c}
\toprule
\textbf{Systematic effect}                              &   \textbf{Affected quantities}    & \textbf{Estimated impact} \\
\midrule
\multirow{1}{*}{Temperature-dependent fiber coupling}   &   Relative light yield              & $2.0\,\%$ \\[6pt]
                                
\multirow{1}{*}{Room temperature fluctuation}           &   Relative / absolute light yield     & $0.1\,\%$ \\[6pt]

\multirow{1}{*}{Internal temperature readout drift}     &   Relative temperature            & $\leq 1.0\,\%$ \\[6pt]        

\multirow{1}{*}{Rotational angular uncertainties}       &   Relative / absolute light yield     & $0.4\,\%$ \\[6pt]                    
\multirow{1}{*}{Spectral calibration filter width}      &   Spectrum                        & $1.0\,$nm \\[6pt]                                
\multirow{1}{*}{TDC time calibration}                   &   Time profile                    & $0.2\,$ns \\[6pt] 
                                                        
\multirow{1}{*}{Picoammeter drift / calibration}    &   Relative / absolute light yield              & $1.0\,\%$ \\[6pt]

\multirow{1}{*}{Self-monitoring sensor aging}    &   Relative / absolute light yield              & $<1.0\,\%$ \\[6pt]

\multirow{1}{*}{Point source approximation}             &   Absolute light yield            & $<1.0\,\%$ \\[6pt]

\multirow{1}{*}{Flashing frequency stability}           &   Absolute light yield            & $<1.0\,\%$ \\[6pt]

\multirow{1}{*}{NIST calibration}                       &   Absolute light yield            & $\leq 0.7\,\%$ \\[6pt]

\multirow{1}{*}{Simulation mismatch}                    &   Absolute light yield            & $\leq 1.0\,\%$ \\

\midrule

\multirow{5}{*}{\textbf{Total}}                         &   Spectrum                        & $1.0\,$nm \\
                                                        &   Time profile                    & $0.2\,$ns \\
                                                        &   Relative temperature            & $1.0\,\%$ \\
                                                        &   Relative light yield            & $2.5\,\%$ \\
                                                        &   Absolute light yield            & $4.1\,\%$ \\
\bottomrule
\end{tabular*}%
\caption{Summary of estimated systematic uncertainties of the POCAM calibration chain. The top part of the table summarizes the dominating systematic effects in the calibration setups, the bottom part shows the total estimated systematic uncertainties on different calibration parameters.}
\label{tab:systematics}
\end{table}\par\noindent

\begin{figure}[h!]
	\centering
	\begin{subfigure}[b]{0.9\textwidth}
	    \centering
	    \hspace*{-1cm}
	    \includegraphics[width=\textwidth]{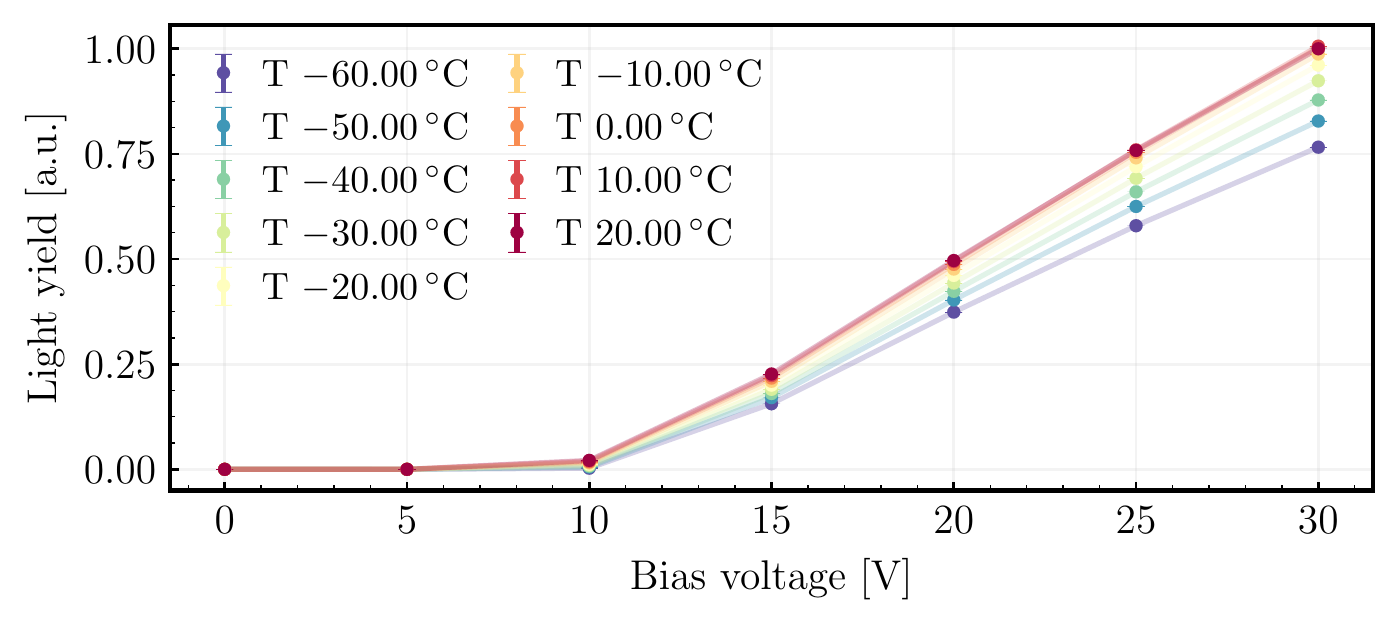}
	    \caption{Default Kapustinsky driver intensity as a function of bias voltage and temperature at $405\,$nm. }
	\end{subfigure}
	\par\vspace{12pt}
	\begin{subfigure}[b]{0.9\textwidth}
	    \centering
	    \hspace*{-1cm}
	    \includegraphics[width=\textwidth]{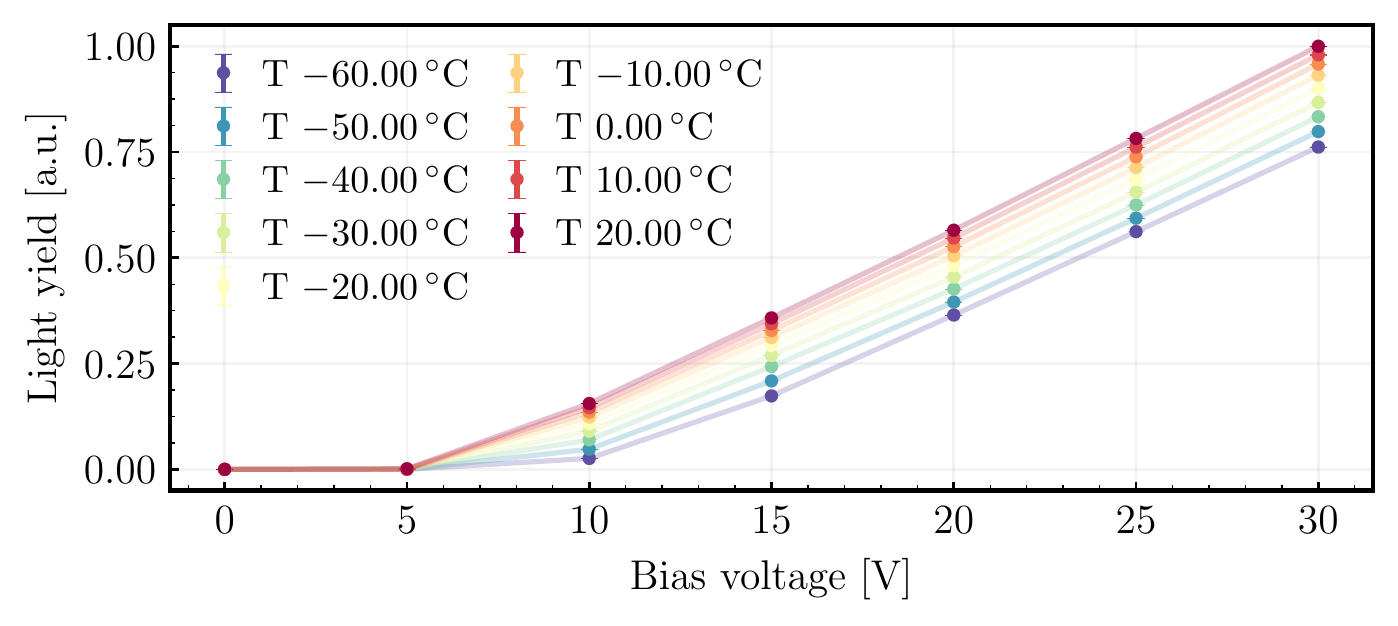}
	    \caption{LD-type (25ns) driver intensity as a function of bias voltage and temperature at $405\,$nm.}
	\end{subfigure}
	\caption{Temperature dependence of the pulser intensity measured for both the default Kapustinsky (a) and LD-type (b) driver using respective $405\,$nm emitters. The fast Kapustinsky shows a similar behavior to the default driver and thus has been omitted in this figure.}
	\label{fig:pulsers-relative}
\end{figure}

\paragraph{Prototype calibration} Preliminary calibration of a full prototype assembly was carried out before going into production. The general procedure first consists of a relative flasher characterization versus configuration parameters and temperature and second a relative emission profile characterization. As for the light emission, shown for the $405\,$nm emitter in \cref{fig:pulsers-relative}, the pulsers show decreasing intensities for decreasing temperatures due to most likely increased series resistance~\cite{ledjunctiontemp}. However, it should be noted that the slope of temperature dependence varies with emitter type. The time profile generally does not show any significant temperature dependence for any driver at higher intensities but the LD-type driver pulses show a marginally longer tail at lower intensities and lower temperatures. However, since all the drivers will be fingerprint-characterized and the LD-type will be predominantly used for high intensities, this does not pose an issue. The emission spectrum showed only small dependence on both temperature and configuration parameters for a small sample of emitters but will be also be characterized for each POCAM.
\par\noindent
The measured emission profile of a POCAM hemisphere prototype and the virtual sum of two hemispheres emulating a complete POCAM, is visualized in Mollweide projection in \cref{fig:moll-single} and \cref{fig:moll-virtual}, respectively. The virtual emission profile shows only marginal deviations from ideal isotropy as was expected from simulation. 
\begin{figure}[h!]
    \centering
    \begin{subfigure}[c]{0.825\textwidth}
        \centering
        \hspace*{0.25cm}
        \includegraphics[width=1\textwidth]{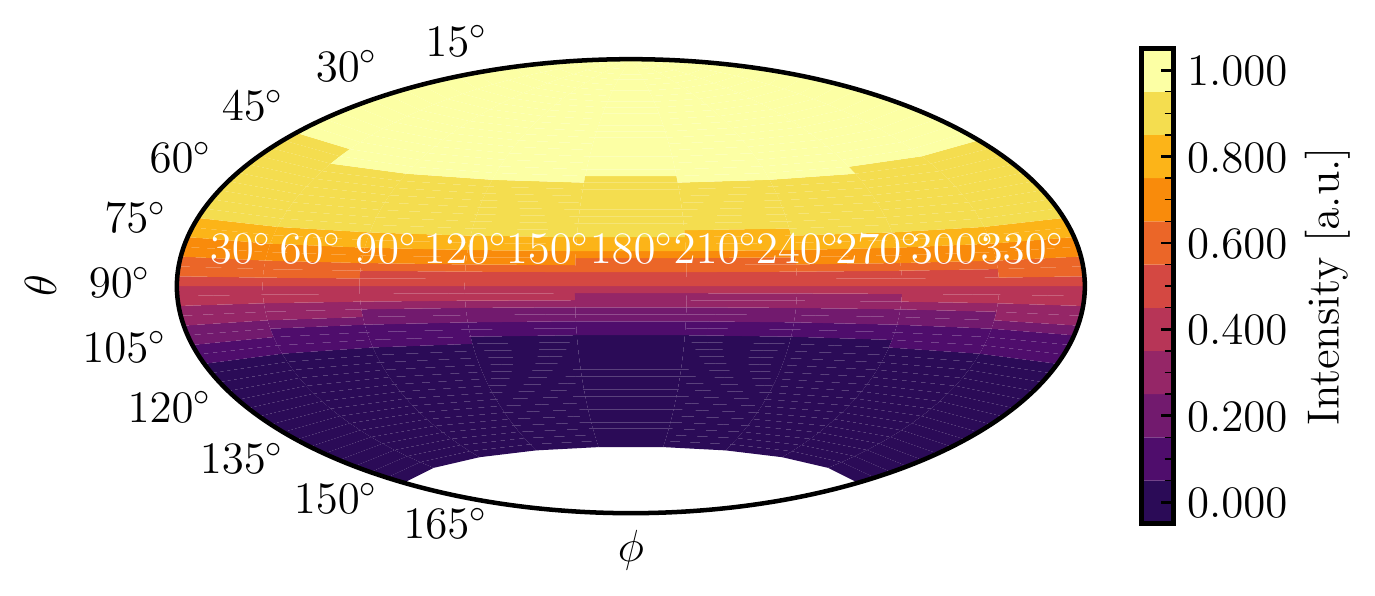}
        \caption{Single hemisphere prototype emission}
        \label{fig:moll-single}
    \end{subfigure}
\end{figure}
\begin{figure}[h!]\ContinuedFloat
    \centering
    \begin{subfigure}[c]{0.825\textwidth}
        \centering
        \hspace*{0.25cm}
        \includegraphics[width=1\textwidth]{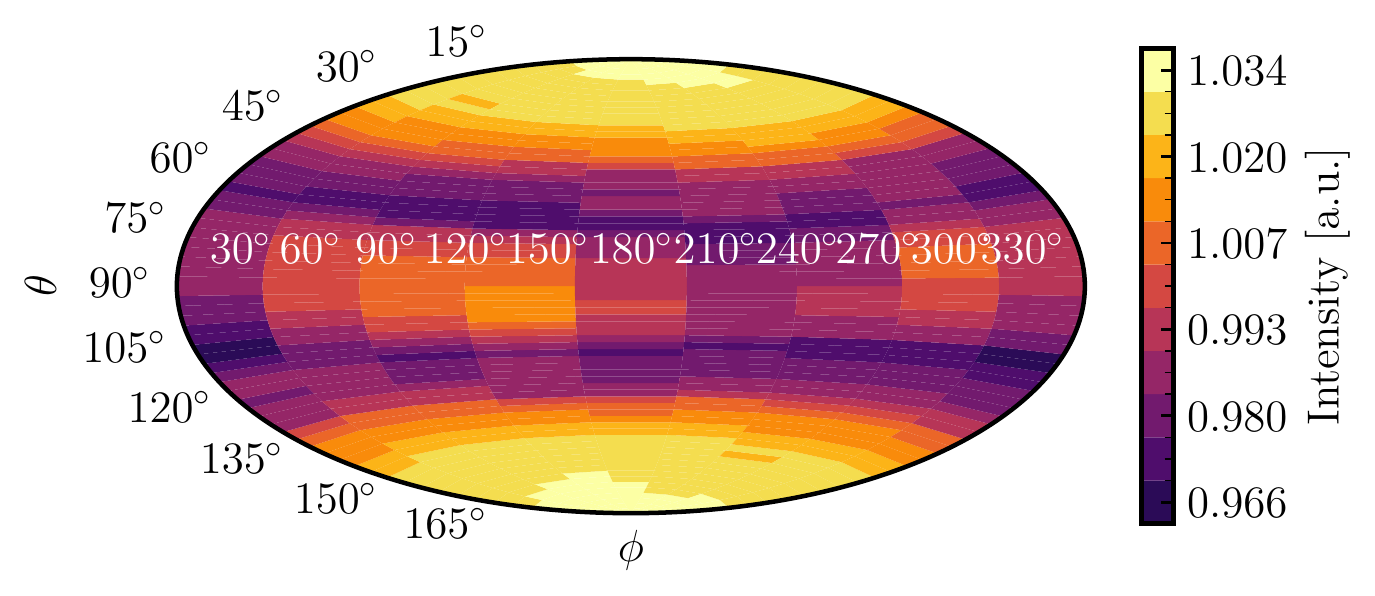}
        \caption{Virtual dual-hemisphere emission}
        \label{fig:moll-virtual}
    \end{subfigure}
    \caption{POCAM hemisphere prototype emission profile in Mollweide projection for a single hemisphere (a) and a virtual complete POCAM (b) in air. In the bottom figure we mirrored and randomly rotated the emission of the hemisphere to create a virtual complete POCAM emission pattern. The pixels represent all measured angular steps with color normalized to maximum (top) and average (bottom) intensity.}
    \label{fig:pocam-angular-moll}
\end{figure}\par\noindent
For its $1\sigma$ deviations we find $0.975 \pm 0.014$ and $1.025 \pm 0.015$ over the full zenith range of $\theta \in [0\degree, 180\degree]$ and $0.986 \pm 0.009$ and $1.014 \pm 0.004$ within $\theta \in [0, 60\degree] \vee [120\degree, 180\degree]$ by averaging all LEDs and all azimuthal angles and using the resulting standard deviation as errors. The hemisphere-to-hemisphere spread is expected to be small due to the precision of production but will nonetheless be characterized. This profile was then further used to roughly estimate the total $4\pi$ POCAM light yield using the $405\,$nm emitter at maximum intensity but yet without NIST-calibration. We find respectively $(5.1 \pm 0.4)\times 10^7\,$photons$\,/\,$pulse for the fast Kapustinsky, $(7.5 \pm 0.6)\times 10^8\,$photons$\,/\,$pulse for the default Kapustinsky and $(2.4 \pm 0.2)\times 10^{10}\,$photons$\,/\,$pulse for the LD-type driver at 25ns width. A dedicated test stand to confirm the $4\pi$ emission profile is currently in planning. 

\section{Summary \& conclusions}
\label{sec:summary}
This work summarizes the developments of the third and final POCAM iteration in the scope of the IceCube Upgrade. In comparison to previous deployments in GVD and STRAW, we have optimized several features of the POCAM including total light yield and subsequent dynamic range, spectral composition of emitters, self-monitoring precision, isotropy and internal structure. Additionally we have developed two dedicated experimental setups which allow a streamlined fingerprint-characterization of individual POCAMs versus temperature, light pulser configuration and orientation. Further we provide an absolute intensity scale using a NIST-calibrated photodiode to provide an absolute scale for all characterization measurements and hence maximize the knowledge of the instrument emission and hence provide what we call golden POCAMs. Lastly, we introduced a modular interface to the detector back end including data stream and synchronization. This can be adjusted to the telescope in which the POCAM is supposed to be deployed in and currently provides an interface to the IceCube detector. With the production of the IceCube instruments starting this year, we aim to provide a large-volume calibration light source standard for large-volume photosensor arrays which can be used in a multitude of scenarios and which provides precise and independent calibration capabilities.

\section*{Acknowledgements}
We would like to express our gratitude to the German Federal Ministry for Education and Research (BMBF) for supporting the development related to IceCube through grant 05A17WO5 "Verbundprojekt IceCube:  Astroteilchenphysik mit dem IceCube-Observatorium". We thank the cluster of excellence "Origin and Structure of the Universe" (DFG) and the SFB1258 "Neutrino and Dark Matter in Astro and Particle Physics" (DFG) for the support related to the GVD and P-ONE applications. We further thank the GVD and STRAW collaborations for deploying our instrument prototypes and sharing of related detector data as well as the IceCube collaboration for comments and feedback on this manuscript.

\bibliography{literature}{}

\begin{thebibliography}{10}

\bibitem{2015RPPh...78l6901A}
Markus {Ahlers} and Francis {Halzen}.
\newblock {High-energy cosmic neutrino puzzle: a review}.
\newblock {\em Reports on Progress in Physics}, 78(12):126901, Dec 2015.

\bibitem{2011A&A...535A.109A}
R.~{Abbasi}, others, and {IceCube Collaboration}.
\newblock {IceCube sensitivity for low-energy neutrinos from nearby
  supernovae}.
\newblock 535:A109, November 2011.

\bibitem{PhysRevLett.114.171101}
A.~Palladino, G.~Pagliaroli, F.~L. Villante, and F.~Vissani.
\newblock What is the flavor of the cosmic neutrinos seen by icecube?
\newblock {\em Phys. Rev. Lett.}, 114:171101, Apr 2015.

\bibitem{Ahlers:2018mkf}
Markus Ahlers, Klaus Helbing, and Carlos Pérez de~los Heros.
\newblock {Probing Particle Physics with IceCube}.
\newblock {\em Eur. Phys. J. C}, 78(11):924, 2018.

\bibitem{reconstruction_icecube}
M.~G. Aartsen et~al.
\newblock {Energy reconstruction methods in the IceCube neutrino telescope}.
\newblock {\em Journal of Instrumentation}, 9(3):P03009, Mar 2014.

\bibitem{Aartsen:2016nxy}
M.~G. Aartsen et~al.
\newblock {The IceCube Neutrino Observatory: Instrumentation and Online
  Systems}.
\newblock {\em JINST}, 12(03):P03012, 2017.

\bibitem{2013Sci...342E...1I}
{IceCube Collaboration}.
\newblock {Evidence for High-Energy Extraterrestrial Neutrinos at the IceCube
  Detector}.
\newblock {\em Science}, 342(6161):1242856, November 2013.

\bibitem{amanda}
E.~{Andres} et~al.
\newblock {The AMANDA neutrino telescope: principle of operation and first
  results}.
\newblock {\em Astroparticle Physics}, 13(1):1--20, Mar 2000.

\bibitem{Collaboration:2011ym}
R.~Abbasi et~al.
\newblock {The Design and Performance of IceCube DeepCore}.
\newblock {\em Astropart. Phys.}, 35:615--624, 2012.

\bibitem{AGERON201111}
M.~Ageron et~al.
\newblock Antares: The first undersea neutrino telescope.
\newblock {\em Nuclear Instruments and Methods in Physics Research Section A:
  Accelerators, Spectrometers, Detectors and Associated Equipment}, 656(1):11
  -- 38, 2011.

\bibitem{Ishihara:2019aao}
Aya Ishihara.
\newblock {The IceCube Upgrade -- Design and Science Goals}.
\newblock In {\em {HAWC Contributions to the 36th International Cosmic Ray
  Conference (ICRC2019)}}, 2019.

\bibitem{2006NIMPA.567..433A}
V.~{Aynutdinov} et~al.
\newblock {The BAIKAL neutrino experiment: From NT200 to NT200+}.
\newblock {\em Nuclear Instruments and Methods in Physics Research A},
  567(2):433--437, Nov 2006.

\bibitem{Margiotta:2014gza}
Annarita Margiotta.
\newblock {The KM3NeT deep-sea neutrino telescope}.
\newblock {\em Nucl. Instrum. Meth.}, A766:83--87, 2014.

\bibitem{2019arXiv190805427B}
{Baikal-GVD Collaboration}.
\newblock {Neutrino Telescope in Lake Baikal: Present and Future}.
\newblock {\em arXiv e-prints}, page arXiv:1908.05427, Aug 2019.

\bibitem{Bedard:2018zml}
M.~Boehmer et~al.
\newblock {STRAW (STRings for Absorption length in Water): pathfinder for a
  neutrino telescope in the deep Pacific Ocean}.
\newblock {\em JINST}, 14(02):P02013, 2019.

\bibitem{2010NIMPA.618..139A}
R.~{Abbasi} et~al.
\newblock {Calibration and characterization of the IceCube photomultiplier
  tube}.
\newblock {\em Nuclear Instruments and Methods in Physics Research A},
  618(1-3):139--152, Jun 2010.

\bibitem{2019PhRvD..99c2007A}
IceCube Collaboration.
\newblock {Measurement of atmospheric tau neutrino appearance with IceCube
  DeepCore}.
\newblock 99(3):032007, Feb 2019.

\bibitem{2013NIMPA.711...73A}
M.~G. {Aartsen} et~al.
\newblock {Measurement of South Pole ice transparency with the IceCube LED
  calibration system}.
\newblock {\em Nuclear Instruments and Methods in Physics Research A},
  711:73--89, May 2013.

\bibitem{Bagley:2009wwa}
P.~Bagley et~al.
\newblock {KM3NeT: Technical Design Report for a Deep-Sea Research
  Infrastructure in the Mediterranean Sea Incorporating a Very Large Volume
  Neutrino Telescope}.
\newblock 2009.

\bibitem{2013ICRC...33.3338C}
Dmitry {Chirkin} and {IceCube Collaboration}.
\newblock {Evidence of optical anisotropy of the South Pole ice}.
\newblock In {\em International Cosmic Ray Conference}, volume~33 of {\em
  International Cosmic Ray Conference}, page 3338, Jan 2013.

\bibitem{Aartsen_2019}
M.G. Aartsen and et~al.
\newblock Measurement of atmospheric tau neutrino appearance with icecube
  deepcore.
\newblock {\em Physical Review D}, 99(3), Feb 2019.

\bibitem{Resconi:2017mad}
Elisa Resconi, Kai Krings, and Martin Rongen.
\newblock {The Precision Optical CAlibration Module for IceCube-Gen2: First
  Prototype}.
\newblock {\em PoS}, ICRC2017:934, 2018.

\bibitem{Aguilar:2004nw}
J.~A. Aguilar et~al.
\newblock {Transmission of light in deep sea water at the site of the ANTARES
  Neutrino Telescope}.
\newblock {\em Astropart. Phys.}, 23:131--155, 2005.

\bibitem{2007NIMPA.578..498A}
M.~{Ageron} et~al.
\newblock {The ANTARES optical beacon system}.
\newblock {\em Nuclear Instruments and Methods in Physics Research A},
  578(3):498--509, Aug 2007.

\bibitem{nanobeacon}
Diego Real and David Calvo.
\newblock Nanobeacon: A time calibration device for km3net.
\newblock {\em EPJ Web of Conferences}, 207:07002, 01 2019.

\bibitem{Balkanov:1999uq}
V.~A. Balkanov et~al.
\newblock {In-situ measurements of optical parameters in Lake Baikal with the
  help of a neutrino telescope}.
\newblock {\em Appl. Opt.}, 33:6818, 1999.

\bibitem{holzapfel:thesis:2019}
K.~Holzapfel.
\newblock { Testing the Precision Optical Calibration Modules in the
  Gigaton-Volume-Detector}.
\newblock Master's thesis, Technical University Munich, 2019.

\bibitem{Fruck:2019vam}
Christian Fruck, Felix Henningsen, and Christian Spannfellner.
\newblock {The POCAM as self-calibrating light source for the IceCube Upgrade}.
\newblock In {\em {HAWC Contributions to the 36th International Cosmic Ray
  Conference (ICRC2019)}}, 2019.

\bibitem{Jurkovic:2016kxn}
M.~Jurkovič et~al.
\newblock {A Precision Optical Calibration Module (POCAM) for IceCube-Gen2}.
\newblock {\em EPJ Web Conf.}, 116:06001, 2016.

\bibitem{Ackermann:2017pja}
M.~Ackermann et~al.
\newblock {The IceCube Neutrino Observatory}.
\newblock In {\em {Proceedings, 35th International Cosmic Ray Conference (ICRC
  2017): Bexco, Busan, Korea, July 12-20, 2017}}, 2017.

\bibitem{Balkanov:2002ni}
V.~Balkanov et~al.
\newblock {Simultaneous measurements of water optical properties by AC9
  transmissometer and ASP-15 Inherent Optical Properties Meter in Lake Baikal}.
\newblock {\em Nucl. Instrum. Meth.}, A298:231--239, 2003.

\bibitem{nbk7}
Schott.
\newblock {\em Schott N-BK7}, 1 2014.
\newblock 517642.251.

\bibitem{epoxy}
3M.
\newblock {\em Scotch-Weld DP 460}, 2003.
\newblock Rev. February.

\bibitem{KAPUSTINSKY1985612}
J.S. Kapustinsky et~al.
\newblock A fast timing light pulser for scintillation detectors.
\newblock {\em Nuclear Instruments and Methods in Physics Research Section A:
  Accelerators, Spectrometers, Detectors and Associated Equipment}, 241(2):612
  -- 613, 1985.

\bibitem{epc9126hc}
EPC.
\newblock {\em Development Board EPC9126/EPC9126HC Quick Start Guide}, 2019.
\newblock Rev. 3.0.

\bibitem{365nm}
Roithner LaserTechnik GmbH.
\newblock {\em XSL-365-5E}, 7 2010.

\bibitem{400nm}
Roithner LaserTechnik GmbH.
\newblock {\em XRL-400-5E}, 9 2010.

\bibitem{ld405nm}
Roithner Lasertechnik GmbH.
\newblock {\em RLT405500MG}, 2016.
\newblock Rev. 1.0.

\bibitem{ld450nm}
Laser Components.
\newblock {\em PL TB450B PRELIMINARY}, 2013.

\bibitem{470nm}
Nichia Corporation.
\newblock {\em Specifications for blue LED}, 3 2012.

\bibitem{ld520nm}
Roithner Lasertechnik GmbH.
\newblock {\em LD-520-50MG}, 2019.

\bibitem{berghof}
Berghof Fluoroplastic~Technology GmbH.
\newblock {Optical PTFE - The reference for light}.

\bibitem{magicblack}
Acktar~Advanced Coatings.
\newblock {MagicBlack}.
\newblock \url{https://www.acm-coatings.de/produkt/magic-black/}.
\newblock [Online; accessed 21$^{\text{st}}$ of January, 2020].

\bibitem{AGOSTINELLI2003250}
S.~Agostinelli et~al.
\newblock Geant4—a simulation toolkit.
\newblock {\em Nuclear Instruments and Methods in Physics Research Section A:
  Accelerators, Spectrometers, Detectors and Associated Equipment}, 506(3):250
  -- 303, 2003.

\bibitem{sipm}
Ketek.
\newblock {\em SiPM - Silicon Photomultiplier PM3315-WB-C0}, 2019.
\newblock Rev. 2019-C.

\bibitem{pd}
Hamamatsu.
\newblock {\em Si photodiodes - S2281 series}, 2015.
\newblock Rev. October.

\bibitem{ad549s}
Analog Devices.
\newblock {\em Ultralow Input Bias Current Operational Amplifier}, 2015.
\newblock Rev. K.

\bibitem{pmid28184106}
A.~L. Migdall and C.~Winnewisser.
\newblock {{L}inearity of a {S}ilicon {P}hotodiode at 30 {M}{H}z and {I}ts
  {E}ffect on {H}eterodyne {M}easurements}.
\newblock {\em J Res Natl Inst Stand Technol}, 96(2):143--146, 1991.

\bibitem{Kotera:2015rha}
Katsushige Kotera, Weonseok Choi, and Tohru Takeshita.
\newblock {Describing the response of saturated SiPMs}.
\newblock 2015.

\bibitem{Rosado:2017ebu}
Jaime Rosado.
\newblock {Performance of SiPMs in the nonlinear region}.
\newblock {\em Nucl. Instrum. Meth.}, A912:39--42, 2018.

\bibitem{Sun:2015pwr}
Yujing Sun and Jelena Maricic.
\newblock {SiPMs characterization and selection for the DUNE far detector
  photon detection system}.
\newblock {\em JINST}, 11(01):C01078, 2016.

\bibitem{2008JInst...310001L}
P.~K. {Lightfoot}, G.~J. {Barker}, K.~{Mavrokoridis}, Y.~A. {Ramachers}, and
  N.~J.~C. {Spooner}.
\newblock {Characterisation of a silicon photomultiplier device for
  applications in liquid argon based neutrino physics and dark matter
  searches}.
\newblock {\em Journal of Instrumentation}, 3(10):001, October 2008.

\bibitem{Eppeldauer:90}
George Eppeldauer.
\newblock Longterm changes of silicon photodiodes and their use for photometric
  standardization.
\newblock {\em Appl. Opt.}, 29(15):2289--2294, May 1990.

\bibitem{Nagai:2019uaz}
Ryo Nagai and Aya Ishihara.
\newblock {Electronics Development for the New Photo-Detectors (PDOM and D-Egg)
  for IceCube-Upgrade}.
\newblock In {\em {HAWC Contributions to the 36th International Cosmic Ray
  Conference (ICRC2019)}}, 2019.

\bibitem{pmt}
Hamamatsu.
\newblock {\em Photomultiplier Tubes R1925, R1925-01}, 2003.
\newblock Rev. February.

\bibitem{apd}
ID Quantique SA.
\newblock {\em ID100 Visible Single-Photon Detector}, 2019.
\newblock Rev. April.

\bibitem{spec}
Hamamatsu.
\newblock {\em Mini-spectrometer Micro series C12880MA}, 2019.
\newblock Rev. June.

\bibitem{ledjunctiontemp}
Edilson Mineiro, F.L.M. Antunes, and Arnaldo Perin.
\newblock Junction temperature estimation for high power light-emitting diodes.
\newblock pages 3030 -- 3035, 07 2007.

\end{thebibliography}
\bibliographystyle{unsrt}

\end{document}